\documentclass[pdflatex,sn-mathphys-num]{sn-jnl}

\usepackage{graphicx}%
\usepackage{multirow}%
\usepackage{amsmath,amssymb,amsfonts}%
\usepackage{amsthm}%
\usepackage{mathrsfs}%
\usepackage[title]{appendix}%
\usepackage{xcolor}%
\usepackage{textcomp}%
\usepackage{manyfoot}%
\usepackage{booktabs}%
\usepackage{algorithm}%
\usepackage{algorithmicx}%
\usepackage{algpseudocode}%
\usepackage{listings}%

\theoremstyle{thmstyleone}%
\theoremstyle{thmstyletwo}%

\theoremstyle{thmstylethree}%

\usepackage{makecell}
\usepackage{bookmark}

\usepackage{titlesec}

\titleformat{\paragraph}[runin]
  {\normalfont\bfseries} 
  {\theparagraph}       
  {1em}                 
  {}                    
  [.]                   

\titlespacing{\paragraph}
  {0pt}                 
  {1.5ex plus .2ex}     
  {0.5em}               

\begin{document}

\title{Physics-based simulation unlocks scaling laws for generalizable MRI reconstruction}


\author*[1]{\fnm{Hong} \sur{Shang}}\email{h.shang@siat.ac.cn}
\author[1,2]{\fnm{Zijian} \sur{Zhan}}
\author[1]{\fnm{Sen} \sur{Jia}}
\author[3]{\fnm{Xiang} \sur{Fan}}
\author[1]{\fnm{Jintao} \sur{Meng}}
\author[1,5]{\fnm{Yang} \sur{Zhou}}
\author[1,2]{\fnm{Xiang} \sur{Zhou}}
\author[1]{\fnm{Tianyu} \sur{He}}
\author[1,5]{\fnm{Xinran} \sur{Chen}}
\author[3]{\fnm{Qiaoling} \sur{Zeng}}
\author[3]{\fnm{Keyan} \sur{Yu}}
\author*[4]{\fnm{Jianhua} \sur{Yao}}\email{jianhua.yao@gmail.com}
\author*[1,5]{\fnm{Dong} \sur{Liang}}\email{dong.liang@siat.ac.cn}

\affil*[1]{\orgdiv{Shenzhen Institutes of Advanced Technology}, \orgname{Chinese Academy of Sciences}, \orgaddress{\city{Shenzhen}, \country{China}}}

\affil[2]{\orgdiv{School of Biomedical Engineering}, \orgname{Shenzhen University Medical School, Shenzhen University}, \orgaddress{\city{Shenzhen}, \country{China}}}

\affil[3]{\orgdiv{Department of Medical Imaging}, \orgname{Peking University Shenzhen Hospital}, \orgaddress{\city{Shenzhen}, \country{China}}}

\affil[4]{\orgdiv{AI for Life Sciences Lab}, \orgname{Tencent}, \orgaddress{\city{Shenzhen}, \country{China}}}

\affil[5]{\orgdiv{State Key Laboratory of Biomedical Imaging Science and System}, \orgname{Chinese Academy of Sciences}, \orgaddress{\city{Shenzhen}, \country{China}}}

\abstract{
Deep-learning-based magnetic resonance imaging (MRI) reconstruction holds great promise for accelerating acquisitions, yet its clinical translation remains bottlenecked by poor generalizability across unseen scanners, populations, anatomies, and protocols. Gathering sufficiently diverse clinical datasets to overcome this bottleneck is hindered by high acquisition costs, privacy restrictions, and institutional biases. Here, we introduce Scalable Inverse Model (SIM), a framework that systematically co-designs physics-simulated training data, model architecture, and learning paradigms around scalability—leveraging compute scaling to overcome historical data bottlenecks and achieve robust zero-shot MRI reconstruction. Across 14 public and in-house datasets—spanning seven international centers, five vendors, field strengths of 0.3-7~T, over ten anatomical regions, more than ten acquisition protocols, downstream quantitative and molecular tasks, and a 150-scan blinded radiologist study—SIM demonstrates robust zero-shot generalizability, recovering fine anatomical details while suppressing aliasing artifacts where conventional deep-learning methods fail. Attention-map analyses confirm that SIM learns physically explainable, dynamically adaptive reconstruction primitives rather than anatomy-specific priors. SIM establishes a scalable, privacy-preserving, and democratized infrastructure for universal MRI reconstruction, providing a blueprint for generalizable AI across inverse problems in medical imaging.
}




\maketitle

\section{Introduction}\label{sec_intro}

\begin{figure*}[h]
\centering
\includegraphics[width=1\textwidth]{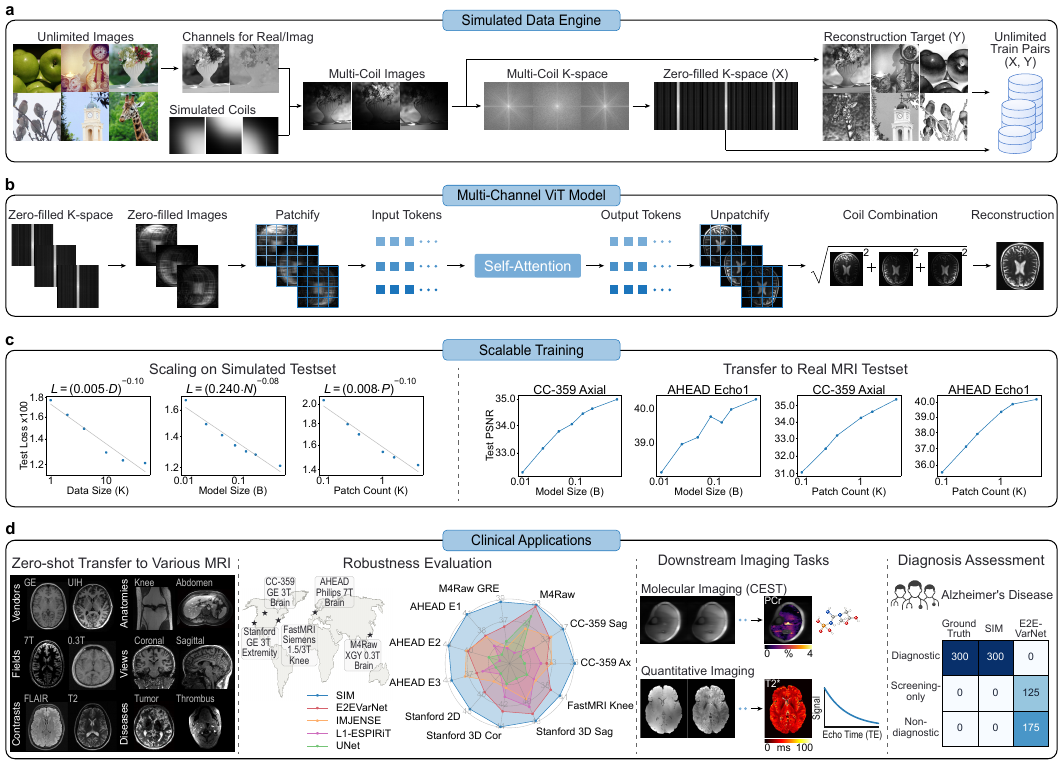}
\caption{Overview of SIM, a compute-scalable framework for zero-shot MRI reconstruction. a, The physics-simulated data engine transforms arbitrary images into paired training data comprising undersampled multi-coil $k$-space measurements ($X$) and ground-truth targets ($Y$). This pipeline strictly obeys multi-coil MRI forward physics to align simulated and real measurement characteristics, effectively closing the simulation-to-real gap. Simultaneously, it expands image and acquisition diversity to overcome historical data bottlenecks, enabling models to learn fundamental inverse mappings. b, SIM adopts the multi-channel Vision Transformer (mcViT)—a high-capacity architecture with minimal MRI-specific inductive bias—to absorb diverse simulated data and learn robust reconstruction priors at scale, replacing hand-crafted architectures designed for data-scarce regimes. c, Pairing large-scale simulated data with high-capacity architecture reveals power-law scaling relationships with dataset size, model size, and patch count, mirroring scaling behaviors in natural language and vision. Crucially, these scaling gains transfer directly to real clinical data, confirming that the simulation engine effectively bridges the simulation-to-real gap. d, Comprehensive zero-shot evaluation across public and in-house datasets—spanning diverse scanner hardware, populations, anatomies, protocols, downstream quantitative and molecular imaging tasks, and a clinical reader study—demonstrates that SIM achieves vastly superior generalizability over traditional real-data learning paradigms.}
\label{fig_overall}
\end{figure*}

Magnetic resonance imaging (MRI) is a cornerstone of clinical diagnosis, offering unparalleled soft-tissue contrast and functional insights without ionizing radiation. Despite decades of advances, MRI remains time-consuming, increasing patient discomfort and motion sensitivity while limiting clinical throughput. Accelerated MRI relies on undersampling measurements in $k$-space—the spatial frequency domain—and reconstructing images using parallel imaging, which exploits multi-coil sensitivities\cite{sense,grappa}, and sparsity driven compressed sensing\cite{mikiCS}. While these approaches form the standard clinical workhorse, further acceleration is fundamentally capped by noise amplification and detail loss. Deep learning holds immense promise for pushing these limits by exploiting data-driven priors to solve ill-posed inverse problems\cite{MoDL,DeepCascade}. Yet, supervised methods often generalize poorly beyond their training distributions\cite{AssessmentSNR,oninstabilities,EvalRobust}. Conversely, untrained methods avoid pretraining by optimizing network parameters on single test cases\cite{ZS-SSL,ConvDecoder,DIPPeder}, but relying strictly on severely undersampled measurements caps their performance\cite{MeasureRobust}. Consequently, the lack of generalizability across diverse imaging conditions remains a central barrier to clinical translation.

Large-scale pretraining on massive datasets has established a paradigm for generalizable AI, from large language models (LLMs)\cite{GPT3} to vision–language models (VLMs)\cite{CLIP}. In medical imaging, foundation models trained on large clinical datasets have improved downstream analysis\cite{HollonNatBME,HollonNatMed,BMEX,MARS,headCTBME2026,xiang2025} and upstream imaging tasks such as CT reconstruction\cite{HorusEye,nict,lungCT}. However, scaling this paradigm in MRI faces substantial data bottlenecks: clinical data collection is costly, datasets may inadequately represent diverse populations and introduce bias\cite{bias2021,bias2023,bias2026}, and privacy and governance constraints impede cross-institutional data sharing\cite{privacy2026,privacyBME2026}. These challenges are particularly acute for MRI reconstruction, which requires raw measurement data rarely retained in clinical archives, and must accommodate a vast range of contrasts and protocols enabled by MRI's flexible parameter configurations.

Synthetic data offer a compelling route to bypass clinical data scarcity. Unlike diagnostic tasks, which often use generative models to synthesize realistic-looking medical images\cite{selfimprove2025,breastgen2026}, inverse problems such as MRI reconstruction can leverage physics-based simulation because the forward encoding process follows well-defined physical laws\cite{bloch1946,jaynes1955,BH2001}. Additionally, non-anatomical physical phantoms have long proven that imaging principles can be established independently of human data\cite{Lauterbur,MRF,MRx}. Building on this principle, prior studies have simulated MRI measurements from existing clinical scans\cite{automap,peder2023,simu_sr2024,simu_tof} or non-medical content, including natural images\cite{AssessmentSNR,PISF}, videos\cite{simu_video} and procedural fractals\cite{simu_fractal}, showing that diverse non-human structures enhance generalization. Simulated data have supported specialized modules for off-resonance correction\cite{resonet} and one-dimensional de-aliasing\cite{PISF}, as well as pretraining\cite{AssessmentSNR,simu_tof} and data augmentation\cite{simu_sr2024}. However, these approaches largely incorporate simulation into conventional architectures, including unrolled networks developed for small-data regimes\cite{VarNet,E2EVarNet}, leaving unclear whether such frameworks fully exploit its potential. Moreover, existing simulation-based methods merely match—rather than surpass—models trained on moderate volumes of real data\cite{peder2023,PISF,simu_video,simu_fractal}. Thus, real-data training remains the dominant paradigm, and a scalable data foundation for generalizable MRI reconstruction remains lacking.

Here, we present Scalable Inverse Model (SIM), a framework centered on the scalable, systematic redesign of data, architecture, and learning paradigms for generalizable MRI reconstruction. For data, SIM leverages physics-based simulation to generate vast, diverse training samples while narrowing the simulation-to-real gap for direct clinical transfer. For architecture, SIM adopts a generic Transformer\cite{ViT} with minimal MRI-specific inductive bias, enabling reconstruction priors to be learned from data at scale rather than explicitly embedded in conventional architectures designed for data-limited regimes. For learning, SIM jointly scales data volume and model capacity, enabling continued performance gains with increased computation and revealing neural scaling laws analogous to those in LLMs\cite{scalinglaw}. Together, these advances establish physics-based simulation not merely as a real-data substitute, but as a foundation for scalable model development. 

Across extensive zero-shot evaluations spanning 14 datasets encompassing seven centers, five vendors, field strengths of 0.3-7~T, more than ten protocols, and diverse anatomies, SIM achieved robust zero-shot reconstruction, consistently outperforming state-of-the-art deep-learning methods. It also generalized to advanced quantitative MRI and molecular MRI without domain-specific training. In a multi-reader study of 150 scans, SIM achieved consistently high diagnostic usability. These findings advance deep-learning reconstruction towards broader clinical translation and establish physics-based simulation as a scalable alternative to clinical data collection, with the potential to alleviate data-sharing constraints and reduce risks associated with privacy and bias. More broadly, this paradigm could extend to other medical imaging inverse problems constrained by limited clinical data. By enabling faster acquisition without compromising diagnostic fidelity, such approaches could improve imaging efficiency, lower hardware demands, and expand healthcare accessibility.

\section{Results}\label{sec_result}
\subsection{Evaluation settings}\label{sec_realdata}
To evaluate generalization, we assembled a comprehensive benchmark comprising 14 datasets from 8 public sources and 6 in-house cohorts, collectively capturing clinically relevant distribution shifts in scanner hardware, populations, anatomies, and protocols (Extended Data Table~\ref{tab_dataset}). All datasets contained multi-coil complex $k$-space data, comprising 12.5 TB of raw data. fastMRI Brain, the largest public dataset acquired at New York University on Siemens scanners using a 2D fast spin-echo (FSE) axial brain protocol\cite{fastMRI}, served as the sole in-distribution (ID) test bed; accordingly, supervised baselines were trained exclusively on its training split. The remaining 13 datasets served as out-of-distribution (OOD) test sets for zero-shot transfer evaluations, spanning ultra-low-field (M4Raw, Xingaoyi, 0.3~T\cite{m4raw}) to ultra-high-field systems (AHEAD, Philips, 7~T\cite{AHEAD}; PCrCEST, UIH, 5~T), diverse international cohorts (Stanford 3D in the US\cite{stanford3d}, M4Raw in China, AHEAD in the Netherlands, and CC-359 in Canada\cite{ccbrain}), and varied anatomical regions (Stanford 2D\cite{stanford2d}; Multi-Anatomy dataset). Retrospective undersampling was applied along the phase-encoding direction ($x$-axis) using 1D equispaced masks at nominal acceleration factors of $4\times$, $6\times$, and $8\times$ with an 8\% fully sampled $k$-space center, yielding effective acceleration factors of $3.2\times$, $4.2\times$, and $5.1\times$, respectively, following fastMRI benchmark protocols\cite{fastMRI}. Comprehensive details regarding dataset preprocessing, ground-truth generation, and evaluation metrics are provided in the Methods.

\subsection{Simulated Data Engine}\label{sec_simudata}

\begin{figure*}[h]
\centering
\includegraphics[width=0.8\textwidth]{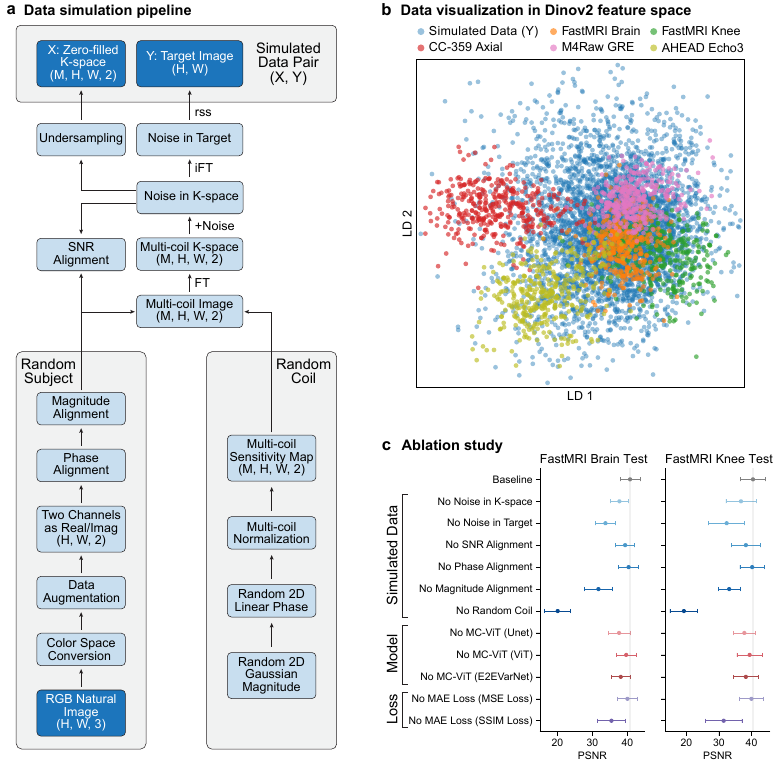}
\caption{Data simulation pipeline, feature representation, and ablation study. a, Data simulation pipeline. The workflow transforms arbitrary natural images into paired training data comprising undersampled multi-coil $k$-space measurements and ground-truth targets. The physical mapping between measurement and target strictly obeys multi-coil MRI forward physics. FT, Fourier transform; iFT, inverse Fourier transform; rss, root-sum-of-squares. b, Data visualization in a shared representation space. While real datasets show a clustered distribution with limited diversity, simulated data display diverse local contrast patterns relevant to clinical imaging at the patch level. Each dot represents a $14\times14$ image patch in the DINOv2 feature space\cite{dinov2} after dimension reduction using principal component analysis and linear discriminant analysis to two primary discriminant axes, color-coded by dataset source. LDA, linear discriminant analysis. c, Ablation study demonstrating the effectiveness of key simulated data design features in closing the simulation-to-real gap, as well as the impact of model architectural choices and learning objectives engineered to leverage simulated training data. Error bars represent mean $\pm$ s.d. Retrospective equispaced undersampling is applied at a $3.2\times$ effective acceleration factor.}
\label{fig_simudata}
\end{figure*}

SIM’s physics-based simulation engine transforms arbitrary spatial patterns from abundant natural images into virtually unlimited paired training data comprising undersampled measurements and ground-truth targets (Fig.~\ref{fig_simudata}a). Training on diverse, non-medical patterns enables the model to learn the underlying inverse mapping while reducing reliance on anatomy-, contrast- and protocol-specific correlations.

A central challenge in simulation-based learning is the simulation-to-real gap. SIM bridges this gap by prioritizing physical measurement fidelity over anatomical similarity through two core strategies. First, we meticulously align the noise characteristics and degradation operators between simulated and real measurements. Undersampling intertwines deterministic aliasing artifacts with stochastic measurement noise\cite{mikiCS}, rendering the reconstruction process highly noise-sensitive\cite{AssessmentSNR,EvalRobust}. To account for substantial noise variations across field strengths and acquisition protocols, SIM spans a broad continuum of signal-to-noise ratios (SNRs). Furthermore, the undersampling operator is explicitly matched between simulation and evaluation, as it defines the forward degradation process whose inverse the reconstruction model must learn.

Second, SIM introduces a dynamic coil simulation strategy to accommodate diverse clinical receiver arrays that cannot be exhaustively predefined. Rather than reproducing a fixed set of coils, we randomly generate a distinct coil configuration for each simulated training example, substantially expanding coil diversity beyond any static clinical dataset, and encouraging generalization to unseen hardware. Ablation experiments validate the contribution of each component (Fig.~\ref{fig_simudata}c). Together, these strategies align simulated and real measurement characteristics while expanding the diversity of image content and acquisition conditions (Fig.~\ref{fig_simudata}b), enabling zero-shot transfer to diverse clinical MRI settings.

\subsection{Scaling laws in reconstruction}\label{sec_scale}

\begin{figure*}[h]
\centering
\includegraphics[width=1\textwidth]{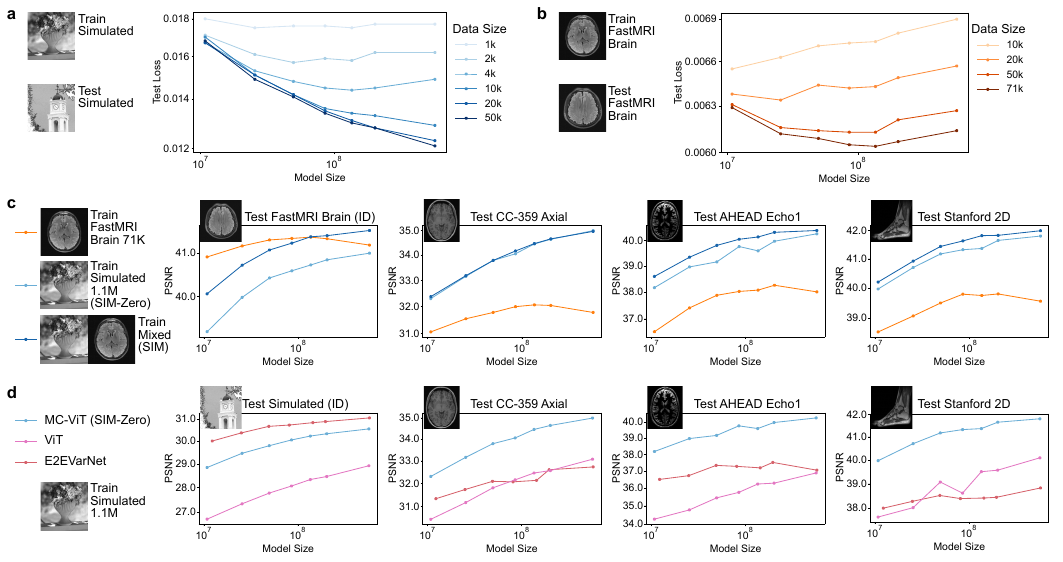}
\caption{Scaling laws in reconstruction. a, In-distribution scaling for simulated data. Reconstruction performance improves smoothly as dataset and model size increase in tandem. Scaling model size yields sustained performance gains when not bottlenecked by data diversity, analogous to scaling laws observed in LLMs\cite{scalinglaw}. b, In-distribution scaling for real fastMRI. Scaling model size fails due to pronounced overfitting (Extended Data Fig.~\ref{ed_scale_loss}b) from inadequate data diversity; despite comparable dataset size to simulated data, real fastMRI exhibits high inter-dataset similarity. c, Transfer scaling comparing training data. Scaling behavior learned on simulated data transfers effectively to diverse real-world datasets, successfully bridging the sim-to-real gap. Real fastMRI alone shows no scaling benefit and transfers poorly to OOD test sets; combining fastMRI with simulated data restores OOD scaling, demonstrating that high-diversity simulated data are essential to unlock scaling laws in MRI reconstruction. d, Transfer scaling comparing model architectures. When trained on 1.1 million simulated samples, conventional architectures designed for small-data regimes with hard-coded domain priors (such as E2EVarNet) are outperformed on OOD test sets by mcViT—a generic Vision Transformer with minimal inductive bias—underscoring the need to co-design data and models for scaling regimes. Experiments in a–c utilize mcViT. Across a–d, retrospective equispaced undersampling is applied at a $3.2\times$ effective acceleration factor. ID, in-distribution.}
\label{fig_scaling}
\end{figure*}

Inspired by the scaling laws observed in LLMs trained on internet-scale data\cite{scalinglaw,GPT3}, we investigated whether MRI reconstruction exhibits analogous scaling behaviour when provided with abundant simulated data. We first examined two scaling dimensions—dataset size and model size—using simulated train and test set. As shown in Fig.~\ref{fig_scaling}a, reconstruction test loss decreased smoothly with increasing dataset and model size in tandem. When dataset size was restricted, increasing model size caused performance to saturate or deteriorate due to overfitting (Extended Data Fig.~\ref{ed_scale_loss}a). Conversely, when model capacity was restricted, increasing the dataset size yielded diminishing returns. Performance followed an approximate power-law relationship with either data or model size when not bottlenecked by the other (Fig.~\ref{fig_overall}c). In contrast, scaling model size with the largest publicly available fastMRI Brain dataset failed to yield sustained improvements (Fig.~\ref{fig_scaling}b), with pronounced overfitting despite a larger training set than some of the simulated-data experiments (Extended Data Fig.~\ref{ed_scale_loss}b). This difference highlights a fundamental limitation of real clinical data: despite their nominal volume, real datasets contain substantial redundancy in anatomy, contrast and acquisition characteristics. The substantially greater diversity of simulated data is essential to unlock scaling laws in MRI reconstruction.

We next examined whether the scaling properties established in simulation transfer to real MRI. We evaluated models up to 567 million parameters under three training regimes: real fastMRI Brain data, 1.1 million simulated samples (SIM-zero), and a combination of both (SIM). For SIM-zero, scaling model size yielded consistent performance gains across diverse real OOD test sets, including CC-359, AHEAD Echo1, and Stanford 2D, without reaching saturation (Fig.~\ref{fig_scaling}c), indicating that the simulation engine effectively bridges the simulation-to-real gap. In contrast, models trained solely on real data exhibited performance degradation beyond modest model size on OOD evaluations. While SIM-zero achieved competitive performance on the ID test set, it substantially outperformed real-data counterpart on OOD cohorts due to its domain-agnostic training. Combining simulated and real data (SIM) further improved performance, particularly on the ID test set, indicating that simulated and clinical data provide complementary learning signals.

Finally, we evaluated how architectural design choices and compute budgets govern scalability. We compared our proposed multi-channel Vision Transformer (mcViT; Methods) against a standard single-channel ViT\cite{MRViT} and E2EVarNet\cite{E2EVarNet}, an unrolled architecture incorporating domain-specific operators such as explicit sensitivity map estimation and data consistency module. When trained across a comparable range of model sizes on 1.1 million simulated samples, all three architectures exhibited scalable performance, but mcViT consistently achieved the superior reconstruction quality (Fig.~\ref{fig_scaling}d). Although E2EVarNet performed competitively under ID testing, its architectural inductive biases constrained its zero-shot transferability. This demonstrates that scalable transformer architectures learn generalizable reconstruction directly from data more effectively than models with hard-coded domain priors. Additionally, scaling the compute budget by reducing spatial patch size—thereby increasing sequence length\cite{ViT,DiT}—yielded consistent performance gains (Fig.~\ref{fig_overall}c, Extended Data Fig.~\ref{ed_scale_ps}). Leveraging these scaling dimensions, we selected 510M and 1B parameter mcViT variants trained with a patch size of 5 (4,096 tokens per channel) on 1.1 million simulated samples for downstream evaluations.

\subsection{Zero-shot reconstruction}\label{sec_bench}

\begin{figure*}[htbp]
\centering
\includegraphics[width=1\textwidth]{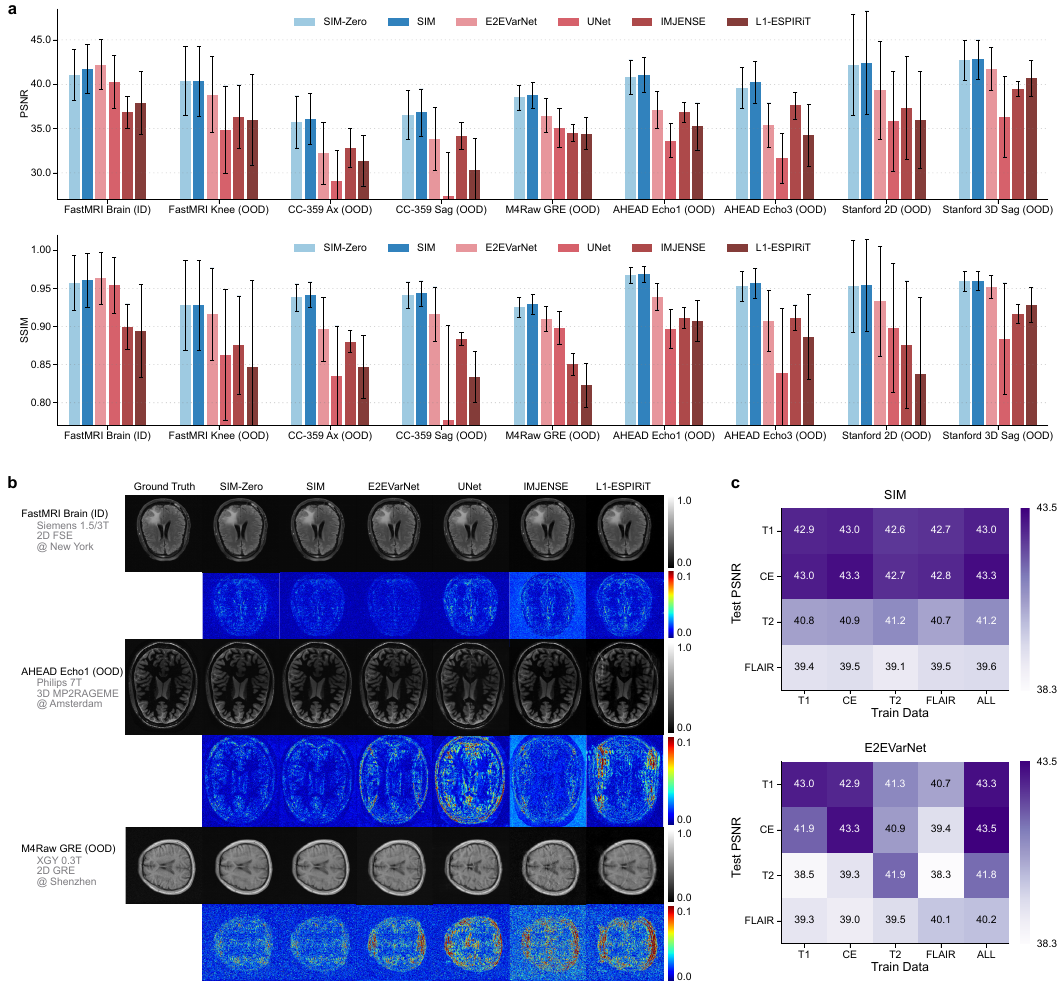}
\caption{Zero-shot reconstruction. a, Quantitative comparison. Both SIM and SIM-Zero (trained exclusively on simulated data) outperform all competing baselines across 8 out-of-distribution (OOD) benchmarks, with SIM exceeding the top baseline by 1.06–3.95 dB in PSNR and 0.008–0.045 in SSIM, and SIM-Zero performing only marginally below SIM. Error bars indicate mean $\pm$ s.d. Full comparisons against 9 state-of-the-art methods are in Extended Data Fig.~\ref{ed_bench_full}. ID, in-distribution. b, Qualitative comparison. Representative reconstructions, error maps, and reference images. SIM recovers fine anatomical detail matching ground truth without perceptible aliasing artifacts (see Extended Data Fig.~\ref{ed_bench_examples} for additional visual comparisons). XGY, Xingaoyi. c, Cross-protocol transfer. Models were trained on fastMRI Brain data partitioned by protocol (T1, contrast-enhanced T1 [CE], T2, FLAIR). E2EVarNet trained on real data exhibits a diagonal performance, degrading on unobserved protocols due to overfitting. Conversely, SIM displays uniform performance across protocols due to domain-agnostic simulated data. Across a–c, retrospective equispaced undersampling is applied at a $3.2\times$ effective acceleration factor.}
\label{fig_bench}
\end{figure*}

To evaluate zero-shot generalization, we retrospectively undersampled diverse public datasets under the data partitioning defined above (Extended Data Table~\ref{tab_dataset}) at an effective $3.2\times$ acceleration factor. We compared SIM with representative state-of-the-art methods spanning major reconstruction paradigms, including end-to-end networks\cite{Unet}, unrolled architectures\cite{E2EVarNet,PDAC,NoVar}, diffusion models\cite{DDS}, untrained methods\cite{IMJENSE,SCAMPI,Aespa} and compressed sensing\cite{ESPIRiT}, with strongest-performing baselines in Fig.~\ref{fig_bench} and complete results in Extended Data Figs. \ref{ed_bench_full} and \ref{ed_bench_examples}. SIM consistently outperformed all competing baselines across every OOD benchmark, exceeding the second-best method by 1.06 to 3.95 dB in PSNR and 0.008 to 0.045 in SSIM (Fig.~\ref{fig_bench}a). Supervised baselines degraded sharply under distribution shifts, with the top-performing model (E2EVarNet\cite{E2EVarNet}) suffering an in-to-out-of-distribution drop of up to 10.01 dB PSNR and 0.067 SSIM. In contrast, SIM demonstrated remarkable stability, capping maximum performance degradation at 5.64 dB PSNR and 0.033 SSIM. IMJENSE\cite{IMJENSE}, the strongest untrained method, sacrificed ID accuracy without providing a consistent zero-shot advantage over E2EVarNet. Notably, SIM-Zero, trained exclusively on simulated data, performed only marginally below SIM and outperformed all baselines across all OOD benchmarks. The performance delta between SIM and SIM-Zero was confined primarily to the ID test set, confirming that incorporating clinical training data provides localized alignment without compromising robustness. Qualitative evaluations of sequential, unselected cases further illustrate this advantage: SIM recovered fine anatomical detail closely matching fully sampled references without perceptible aliasing artifacts, whereas competing methods exhibited severe aliasing and distortions under distribution shifts, such as ultra-low-field of 0.3~T\cite{m4raw} and ultra-high-field of 7~T\cite{AHEAD} (Fig.~\ref{fig_bench}b and Extended Data Fig.~\ref{ed_bench_examples}). Furthermore, unlike diffusion and untrained approaches that require hundreds of iterations per sample, SIM operates via a single feed-forward inference step. This yields low latency (Extended Data Fig.~\ref{ed_bench_full}b) suitable for clinical integration while delivering unmatched robustness across diverse scanner hardware, populations, and protocols.

We further evaluated SIM across an expanded spectrum of anatomical regions and clinical contrasts using public and in-house data at an effective acceleration of $4.2\times$. As shown in Extended Data Fig.~\ref{ed_anat_cont}a, SIM demonstrated robust performance across 11 distinct anatomical structures, including five anatomical acquisitions from our in-house Multi-Anatomy dataset (cervical spine, thoracic spine, lumbar spine, pelvis, and heart) and six public scans from the Stanford 2D dataset\cite{stanford2d} (shoulder, elbow, hand, ankle, thigh, and foot). Across these anatomies, SIM closely matched the fully sampled references without perceptible artifacts or blurring. Furthermore, SIM consistently maintained high fidelity across diverse clinical contrasts in our in-house Multi-Contrast cohort (neck protocol with three contrasts; brain protocol with four contrasts) and an in-house dynamic contrast-enhanced (DCE) liver cohort (seven dynamic phases), covering patients with thrombi, brain tumor, and hepatic lesion respectively (Extended Data Fig.~\ref{ed_anat_cont}b). Across these varied contrasts and disease presentations, SIM preserved critical lesion morphology and subtle tissue boundaries without introducing undersampling artifacts, confirming its broad applicability across routine clinical protocols.

To investigate the mechanisms driving protocol transfer, we partitioned the fastMRI Brain dataset\cite{fastMRI} into four protocol-specific subsets ($T_1$, contrast-enhanced $T_1$, $T_2$, and $T_2$-FLAIR) and trained separate models on each to evaluate cross-protocol transfer (Fig.~\ref{fig_bench}c). While SIM and E2EVarNet achieved comparable performance when trained on the entire dataset, their transfer behaviours differed markedly when training were restricted to individual protocols. E2EVarNet exhibited a diagonal pattern, performing well on matching tests but degrading on unobserved contrasts, while SIM showed a more uniform pattern across training protocols. These findings confirm that real data training induces protocol overfitting, whereas SIM achieves cross-protocol invariance due to its domain-agnostic training on non-medical content.

\clearpage

\subsection{Zero-shot downstream applications}\label{sec_down}
Quantitative and molecular MRI provide quantitative tissue properties and molecular biomarkers unattainable through routine protocols, enabling precise tissue characterization and enhanced disease subtyping\cite{MRx,qmrireview2021}. However, clinical translation remains bottlenecked by prolonged acquisitions—whether from multi-contrast sampling in quantitative MRI\cite{qmrireview2023} or molecular techniques like chemical exchange saturation transfer (CEST) MRI, which demand rapid readouts to maintain subtle saturation-transfer contrast\cite{cestreview2022}. Accelerating these emerging sequences using supervised models is challenging due to the lack of curated clinical training datasets. SIM bypasses this bottleneck through robust zero-shot reconstruction. Retaining fastMRI Brain as the sole real dataset during training, we evaluated SIM's zero-shot acceleration under retrospective undersampling across two advanced tasks: quantitative $T_2^*$ mapping and dynamic molecular phosphocreatine/creatine CEST (PCrCEST/CrCEST) imaging.

\paragraph{Quantitative MRI}\label{sec_qmri}

\begin{figure}[h]
\centering
\includegraphics[width=0.8\textwidth]{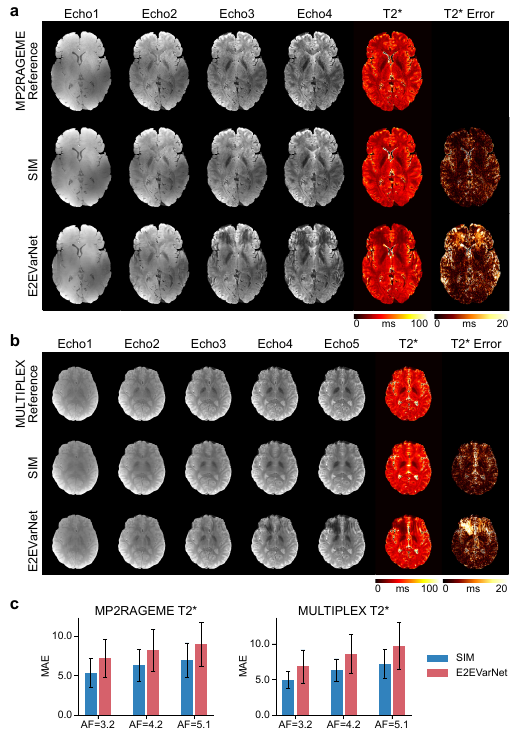}
\caption{Zero-shot quantitative MRI. a,b, Representative single-echo reconstructions, derived $T_2^*$ maps, and corresponding error maps evaluated on the public AHEAD dataset (MP2RAGEME sequence, Philips 7~T) (a) and an in-house cohort (MULTIPLEX sequence, UIH 3~T) (b). SIM's zero-shot capability enables reliable downstream parametric mapping. Skull stripping is applied, and display windowing and leveling are adjusted per echo image and matched between reconstructions and reference images. Retrospective equispaced undersampling is applied at a $3.2\times$ effective acceleration factor. c, Quantitative comparison of $T_2^*$ mapping across all slices evaluated by mean absolute error (MAE) at effective acceleration factors of $3.2\times$, $4.2\times$, and $5.1\times$. SIM consistently outperforms baselines across sequences and acceleration factors. Error bars represent mean $\pm$ s.d. AF, acceleration factor.}
\label{fig_qmr}
\end{figure}

We first evaluated representative $T_2^*$ mapping using two multi-parametric sequences, MP2RAGEME\cite{mp2rageme} (four echoes of the public AHEAD dataset\cite{AHEAD}) and MULTIPLEX\cite{mtp} (five echoes of our in-house MULTIPLEX dataset) (Extended Data Table~\ref{tab_dataset}). SIM and E2EVarNet were used to reconstruct each echo image independently, at effective acceleration factors of $3.2\times$, $4.2\times$, and $5.1\times$, followed by a standardized mono-exponential decay fitting. For both sequences as shown in Fig.~\ref{fig_qmr}a,b, SIM achieved high-fidelity reconstruction across individual echo image, propagating accurate signal decay dynamics into the resulting $T_2^*$ maps relative to fully sampled references. In contrast, E2EVarNet’s echo-level aliasing, distortions and signal loss compounded during fitting into spurious spatial variations across resulting $T_2^*$ maps that compromise diagnostic utility. Quantitative evaluation confirmed this advantage (Fig.~\ref{fig_qmr}c): across all acceleration factors and sequences, SIM reduced mean absolute error (MAE) by at least 1.83 ms relative to E2EVarNet, confirming that SIM’s zero-shot capability ensures reliable downstream parametric mapping.

\clearpage

\paragraph{Molecular MRI}\label{sec_cest}

\begin{figure}[h]
\centering
\includegraphics[width=0.8\textwidth]{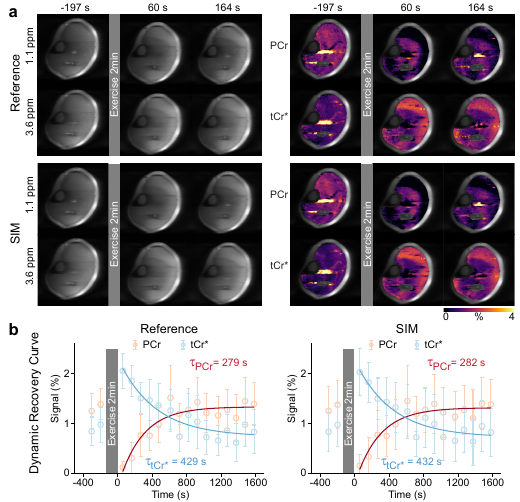}
\caption{Zero-shot molecular MRI. a, Representative SIM reconstructions of dynamic CEST images (shown at offset frequencies of 1.1 ppm and 3.6 ppm) and derived dynamic maps of PCr and $t\text{Cr}^*$ (a weighted sum of PCr and Cr\cite{pcrcest}), evaluated on an in-house dataset monitoring calf-muscle recovery using an in-scanner exercise protocol\cite{pcrcest} (UIH 5~T). b, Dynamic recovery curves within the calf-muscle region of interest (ROI). SIM achieves image quality virtually identical to the reference across all dynamic timepoints, enabling faithful preservation of dynamic physiological responses. For b, data points represent mean values, and error bars indicate s.d. across all pixels within the calf-muscle ROI. For a and b, retrospective equispaced undersampling is applied at a $5.1\times$ effective acceleration factor. PCr, phosphocreatine; Cr, creatine; $\tau$, metabolic recovery time constant.}
\label{fig_cest}
\end{figure}

We next evaluated SIM for dynamic PCrCEST/CrCEST imaging using an in-house dataset acquired at UIH 5~T scanner to monitor calf-muscle recovery following an in-scanner exercise protocol\cite{pcrcest}. In brief, a subject in the supine position performed a single-leg plantar-flexion exercise for 120 s, with two pre-exercise and 15 post-exercise acquisitions at approximately 100 s intervals. SIM was applied to reconstruct individual CEST image at a $5.1\times$ effective acceleration factor followed by the same processing pipeline as in\cite{pcrcest}. SIM achieved virtually identical CEST image quality to fully sampled references across all dynamic timepoints (Fig.~\ref{fig_cest}a). This yielded spatial maps of PCr and $t\text{Cr}^*$ (a weighted sum of PCr and Cr\cite{pcrcest}) that closely matched reference maps (Fig.~\ref{fig_cest}a), leading to highly accurate kinetic recovery curves within the calf-muscle region of interest (Fig.~\ref{fig_cest}b). Crucially, SIM derived accurate metabolic recovery time constants—PCr (282~s vs 279~s reference) and $t\text{Cr}^*$ (432~s vs 429~s reference)—demonstrating faithful preservation of dynamic physiological responses.

\subsection{Zero-shot clinical applications}\label{sec_reader}

\begin{figure*}[h]
\centering
\includegraphics[width=1\textwidth]{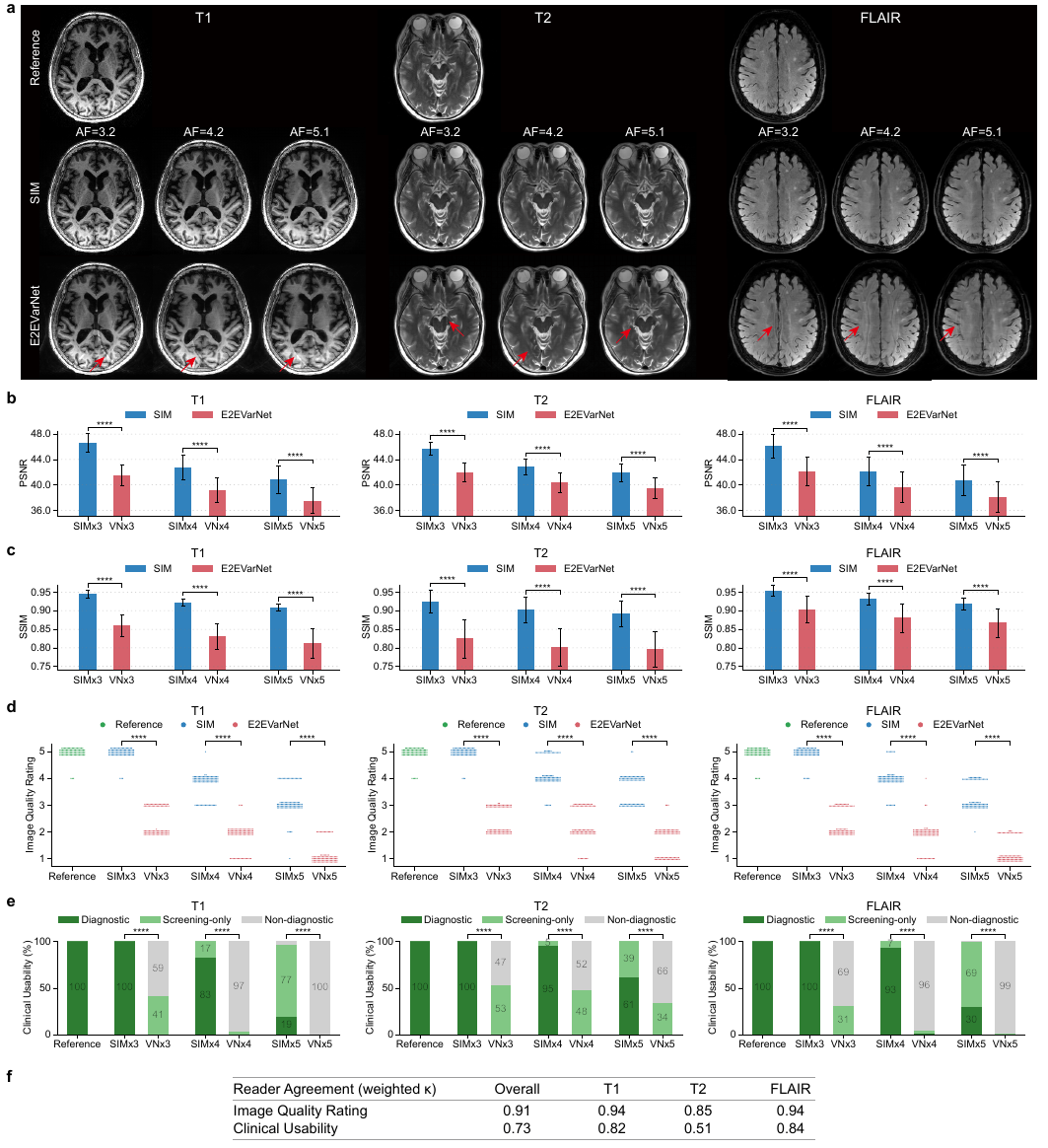}
\caption{Zero-shot clinical applications. a, Representative reconstructions of SIM and E2EVarNet at effective acceleration factors of $3.2\times$, $4.2\times$, and $5.1\times$ for an in-house Alzheimer’s disease cohort with three contrasts: $T_1$-weighted, $T_2$-weighted, and $T_2$-FLAIR. Red arrows indicate residual aliasing artifacts. b,c, Quantitative evaluation in terms of PSNR (b) and SSIM (c). Error bars represent mean $\pm$ s.d. d–f, Blinded reader study in which 1,050 MRI volumes were evaluated independently by two board-certified radiologists. d, Radiologist ratings of overall image quality evaluated on a 5-point Likert scale ($5 = \text{excellent}$, $1 = \text{unacceptable}$); each dot represents an individual rating. e, Radiologist ratings of clinical usability evaluated on a 3-point categorical scale (diagnostic, screening-only, non-diagnostic). f, Inter-reader agreement for image quality and clinical usability ratings. Statistical significance was assessed using two-sided paired Student’s $t$-tests for quantitative metrics (b,c) and two-sided clustered Wilcoxon signed-rank tests for radiologist ratings (d,e); significance levels are denoted as ****$P < 0.0001$. Inter-reader agreement was quantified using quadratically weighted Cohen’s $\kappa$ for image quality ratings and linearly weighted Cohen’s $\kappa$ for clinical usability. For d and e, ratings from two radiologists are pooled for visualization. AF, acceleration factor; VN, E2EVarNet; x3, x4, and x5, effective acceleration factors of $3.2\times$, $4.2\times$, and $5.1\times$, respectively.}
\label{fig_ad}
\end{figure*}

To evaluate zero-shot reconstruction in a clinical context, we conducted a blinded reader study on an in-house Alzheimer’s disease cohort (STAR; Peking University Shenzhen Hospital; 3~T UIH). The dataset comprised 50 subjects (25 healthy controls, 25 patients) without evident motion artefacts, acquired using a standard brain protocol\cite{adni4}: 3D MPRAGE $T_1$-weighted, 2D FSE $T_2$-weighted, and $T_2$-FLAIR contrasts. Zero-shot performance of SIM and E2EVarNet was evaluated using retrospective undersampling at effective acceleration factors of $3.2\times$, $4.2\times$, and $5.1\times$. For each subject and contrast, seven images were assessed: the fully sampled reference and reconstructions from both methods at the three acceleration factors, yielding 1,050 volumes in total. Two board-certified radiologists ($>10$~years of experience) independently evaluated the randomized, anonymized volumes presented alongside the fully sampled reference. Reconstructions of the same subject were shuffled with a minimum 15-case washout interval to minimize recall bias. Image quality was rated on a 5-point scale (5 = excellent, 1 = unacceptable), and clinical usability was classified into three ordinal categories: diagnostic, screening-only and non-diagnostic. Ratings from both radiologists were aggregated across methods for display, with detailed statistical procedures provided in the Methods.

Visual inspection confirmed that SIM closely reproduced the reference images at $3.2\times$ acceleration and maintained alias-free quality across higher acceleration levels for all contrasts, whereas E2EVarNet exhibited pervasive aliasing artifacts (Fig.~\ref{fig_ad}a). Quantitatively, SIM significantly outperformed E2EVarNet across all contrasts and acceleration factors ($P < 0.0001$) in PSNR, SSIM, radiologist-rated image quality, and clinical usability (Fig.~\ref{fig_ad}b-e), backed by strong inter-reader agreement (weighted Cohen’s $\kappa = 0.91$ for image quality; $\kappa = 0.73$ for clinical usability; Fig.~\ref{fig_ad}f). Crucially, at $3.2\times$ acceleration, SIM achieved 100\% diagnostic usability, matching the fully sampled reference, and received top image-quality score (5/5) in 95–97\% of ratings across contrasts. Conversely, E2EVarNet at $3.2\times$ fell into screening-only or non-diagnostic categories, and never scored above 3/5 for image quality. Although image quality degraded for both methods at higher acceleration rates, SIM consistently maintained clinical utility with far fewer non-diagnostic failures; at $5.1\times$ acceleration, non-diagnostic rates for E2EVarNet were 100\% ($T_1$), 66\% ($T_2$), and 99\% ($T_2$-FLAIR), compared to only 4\%, 0\%, and 1\% for SIM. Notably, expert ratings revealed a far steeper clinical penalty for E2EVarNet than pixel-wise metrics (PSNR, SSIM), underscoring that localized artifacts—which are smoothed out by global pixel averages—are clinically intolerable due to the risk of pseudo-lesion hallucinations. Together, these results demonstrate that SIM’s generalizability translates into consistently high clinical usability, with aliasing-free, high-quality zero-shot reconstructions that were not achieved by the real-data-trained baseline.

\clearpage

\subsection{Interpretability of zero-shot generalization}\label{sec_explain}

\begin{figure*}[h]
\centering
\includegraphics[width=1\textwidth]{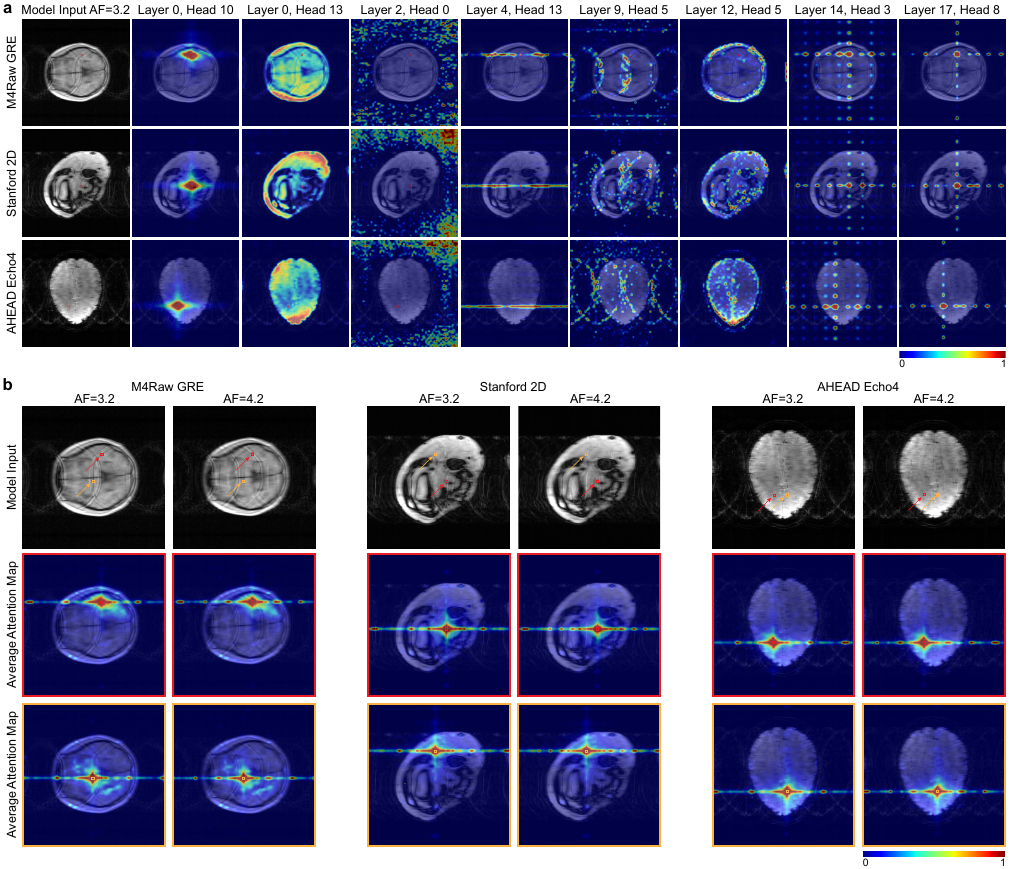}
\caption{Visualization of SIM attention maps. a, Attention maps of individual heads across layers, demonstrating that SIM intrinsically captures generic image structures and spatial relationships tied to undersampling physics via large-scale domain-agnostic training, thereby achieving physically explainable and broadly generalizable reconstruction. b, Attention maps aggregated across all layers and heads, showing that SIM dynamically adapts its attention to aliasing source regions as changes in acceleration alter the spatial aliasing distance. Two example query patches are shown under effective acceleration factors of $3.2\times$ and $4.2\times$. For a and b, arrows and boxes on input images indicate query patches. AF, acceleration factor.}
\label{fig_attnmap}
\end{figure*}

To investigate the mechanism underlying SIM’s zero-shot generalization, we visualized the attention maps of its Transformer backbone to probe internal representations during reconstruction. For a given query patch, these attention maps highlight the spatial regions receiving the highest attention weights. Fig.~\ref{fig_attnmap} presents three examples across diverse acquisition settings, including ultra-low-field (M4Raw GRE, 0.3~T), ultra-high-field (AHEAD Echo4, 7~T), and distinct anatomical regions (brain and knee). Visual inspection revealed that individual heads across layers specialized in distinct, semantically meaningful, and reproducible patterns, including local neighborhoods, foreground, background, structural boundaries, and acceleration-dependent axial aliasing signatures (Fig.~\ref{fig_attnmap}a). These patterns indicate that SIM captures spatial relationships intrinsically tied to $k$-space undersampling physics and generic image structure, rather than relying on anatomy- or contrast-specific priors. Training on vast non-medical simulated data prevents overfitting, forcing the model to learn broadly applicable, physically explainable representations that transfer robustly to real clinical acquisitions. Furthermore, layer-aggregated attention maps confirm that SIM resolves undersampling artifacts by attending precisely to their originating source regions, directly aligning with the physical relationships imposed by undersampling (Fig.~\ref{fig_attnmap}b). Crucially, SIM dynamically adapts its attention when changes in acceleration alter the spatial aliasing distance. For instance, in an M4Raw GRE acquisition, aliasing for a target patch (outlined in red) originated from the anterior skull at $3.2\times$ acceleration but shifted to the posterior skull at $4.2\times$; SIM’s attention shifted to the respective source region in each case (Fig.~\ref{fig_attnmap}b). Together, these findings confirm that SIM’s zero-shot generalization across unseen clinical domains stems from learning physically explainable, dynamically adaptive representations.

\section{Discussion}\label{sec_discuss}
In this study, we present SIM, a scalable framework trained on physics-simulated data to enable zero-shot accelerated MRI across highly diverse clinical environments. Extensive validation across public and in-house datasets—spanning varied scanner hardware, populations, anatomies, protocols, downstream quantitative/molecular MRI applications, and a clinical reader study—demonstrates that SIM achieves robust generalizability, substantially surpassing conventional deep-learning methods. A core driver of SIM’s performance is its shift from isolated data synthesis or specialized model design toward the co-design of data, architecture, and learning around scalability. While conventional architectures embedded hand-crafted priors for small-data regimes, they bottleneck the expressivity needed for massive diversity of simulated data. By pairing a physics-faithful simulation engine with a high-capacity generic architecture, SIM converts increasing compute into continuous performance gains—mirroring scaling laws in language\cite{scalinglaw} and vision\cite{DiT}. Consequently, SIM proves that diverse simulated data, when fully exploited within a scalable framework, yields vastly superior generalizability compared to traditional real-data training paradigms.

Beyond improving generalizability, physics-simulated data addresses several structural limitations of real clinical datasets—including scalability bottlenecks\cite{HollonNatBME,HollonNatMed}, algorithmic bias\cite{bias2021,bias2023,bias2026}, patient privacy risks\cite{privacy2026,privacyBME2026}, and regulatory barriers to data sharing\cite{breastgen2026}. While collecting large multi-center clinical cohorts is prohibitively expensive and inherently fragmented, simulation engines operate on invariant physical laws rather than variable human anatomy and pathology, offering near-zero marginal cost and unlimited scaling. Training on non-medical simulated data also mitigates dependence on clinical datasets that often encode geographic, demographic, and institutional biases, establishing a more equitable foundation, though the precise extent to which this enhances fairness warrants systematic evaluation. Finally, distributing open-source simulation code instead of protected health data bypasses patient privacy concerns and regulatory barriers—establishing a scalable, decentralized data infrastructure for reproducible, collaborative AI development.

While demonstrated here on accelerated MRI, the rationale behind SIM applies broadly across inverse problems in medical imaging. Tasks such as MRI denoising\cite{denoise2026} and super-resolution\cite{BMEX} for portable MRI\cite{lfscience2024,lfnc2024}, as well as low-dose or limited-angle computed tomography\cite{HorusEye,nict}, and low-count positron emission tomography\cite{PET2021,PET2026}, share two principles: the objects being imaged need not be human anatomy, and the acquisition process can be simulated using well-established physics. Vast, non-medical content can supply the structural diversity needed for model generalization, while physics-based simulation makes generating such massive datasets practical and scalable. By combining these principles with scalable model training, SIM establishes a generalizable blueprint for modalities long constrained by clinical data scarcity. Ultimately, democratizing zero-shot inverse models could lower hardware specifications, shorten acquisition times, and reduce radiation exposure without compromising diagnostic fidelity, improving healthcare accessibility and affordability.

Several technical and clinical avenues remain for further advancement. First, our study focused on two-dimensional multi-slice, magnitude-only MRI to establish a foundational proof-of-concept. However, three-dimensional (3D) isotropic MRI and phase-sensitive imaging\cite{phasemri2020} are increasingly vital in clinical practice, with 3D acquisitions accounting for a substantial proportion of total scan time\cite{mrireview2024,mrireview2025}. Extending the physics-based simulation engine to support an additional spatial dimension (3D) while expanding the network’s output channel dimension to accommodate complex-valued reconstructions could enable substantially higher acceleration factors and broader clinical utility. Second, our validation relied on retrospective undersampling to rigorously benchmark cross-domain robustness. Translation to clinical practice will require prospective, multicenter studies to evaluate integration with clinical scanners and establish clinically appropriate acceleration factors. Finally, while we leveraged a generic Transformer architecture for its superior scalability, integrating domain-specific physical priors—like inter-coil correlations—without compromising scalability provides a clear pathway to enhance reconstruction fidelity in future work.

\section{Methods}\label{sec_method}

\subsection{Simulated training data}\label{sec_simudetail}
SIM simulates the forward spatial encoding process of multi-coil MRI based on well-defined physical principles. In multi-coil acquisitions, each receiver coil measures spatial frequency components of the target volume weighted by its complex-valued, position-dependent coil sensitivity map. The $k$-space signal acquired by the $i$-th coil, $\mathbf{k}_i$, is defined as:
\begin{equation}
  \mathbf{k}_i = \mathcal{F}(\mathbf{S}_i \mathbf{m}) + \boldsymbol{\varepsilon}_i, \quad i = 1, 2, \ldots, N
  \label{eq:fullsamp}
\end{equation}
where $\mathbf{m}$ is the 2D imaging object, $\mathbf{S}_i$ is the spatial sensitivity map of the $i$-th receiver coil, $\mathcal{F}$ is the 2D Fourier transform operator, $\boldsymbol{\varepsilon}_i$ is the measurement noise for the $i$-th coil, and $N$ is the total number of receiver coils.

In undersampled MRI, only part of the $k$-space is acquired to accelerate scan
\begin{equation}
  \tilde{\mathbf{k}}_i = \mathcal{M} \mathcal{F}(\mathbf{S}_i \mathbf{m}) + \boldsymbol{\varepsilon}_i, \quad i = 1, 2, \ldots, N
  \label{eq:forwardmodel}
\end{equation}
where $\tilde{\mathbf{k}}_i$ is the undersampled $k$-space data acquired by the $i$-th coil, $\mathcal{M}$ is the binary undersampling mask.

Grounded in the physical formulation above, SIM’s simulation engine incorporates two central design considerations: high measurement fidelity to close the simulation-to-real gap and maximum structural diversity via non-medical patterns. By preventing the model from memorizing anatomical, contrast, or protocol-specific priors, this domain-agnostic approach encourages the learning of fundamental inverse mappings. Overall, the engine transforms arbitrary natural images into paired training data—comprising undersampled $k$-space measurements and ground-truth targets (Fig.~\ref{fig_simudata}a)—via three core modules: random subject generation, random coil simulation, and multi-coil pair synthesis.

For random subject generation, we sample high-resolution natural images from the open-source SA-1B dataset\cite{SAM}. To simulate complex-valued MRI signals with highly diverse phase patterns, we randomly assign two distinct image channels—extracted either from the native RGB color space or converted spaces such as HSV or YCbCr—to represent the real and imaginary components. The image undergoes standard data augmentations to further enrich spatial diversity, including random resizing, cropping, horizontal and vertical flipping, and channel-wise rescaling. Because the initial real and imaginary channels are non-negative, their phase values are inherently restricted to the range $[0, \pi/2]$. We apply a Phase Alignment module to project the phase across the full $[0, 2\pi]$ range. Finally, a Magnitude Alignment module applies scaling factors sampled from a log-normal distribution to diversify the overall signal magnitude across training samples.

For random coil simulation, SIM dynamically generates a distinct coil configuration for each simulated training sample to address the vast diversity of clinical receiver configurations. Rather than reproducing specific hardware layouts, this design maximizes coil profile diversity to foster hardware-agnostic generalization. Specifically, for an array with $N$ coils, we generate $N$ random 2D Gaussian kernels to model sensitivity magnitudes and $N$ 2D linear functions with random coefficients to model phase profiles. Combining these yields $N$ complex-valued coil profiles, which are normalized using pixel-wise root-sum-of-squares (RSS) to produce the final coil sensitivity maps. Given an initial simulated sensitivity map $\mathbf{S}_i$ for the $i$-th coil, the normalized sensitivity map $\tilde{\mathbf{S}}_i$ is defined as:
\begin{equation}
  \tilde{\mathbf{S}}_i = \frac{\mathbf{S}_i}{\sqrt{\sum_{j=1}^{N} |\mathbf{S}_j|^2}}
  \label{eq:csm_norm}
\end{equation}

To synthesize multi-coil $k$-space–target pairs according to Equation~\ref{eq:forwardmodel}, we multiply the complex-valued random subject by the complex-valued coil sensitivity maps to generate multi-coil complex-valued spatial images, which are subsequently transformed into fully sampled multi-coil $k$-space via a 2D fast Fourier transform. Complex additive Gaussian noise is then injected into the $k$-space measurements. Combined with the aforementioned magnitude alignment, we carefully calibrate the signal-to-noise ratio (SNR) across a broad spectrum to encapsulate the noise variations encountered across different field strengths and acquisition protocols. Finally, an undersampling mask $\mathcal{M}$ is applied to yield the zero-filled, undersampled $k$-space data, $\mathbf{X}$. Crucially, the mask distribution during training must encompass the undersampling geometries used at evaluation. When targeting multiple acceleration factors, all corresponding masks are randomly selected for each training sample; excluding specific mask configurations during training causes marked performance degradation when evaluated on those omitted masks (Extended Data Fig.~\ref{ed_multiR}).

The ground-truth target image is generated from the fully sampled, noise-corrupted $k$-space following standard multi-coil MRI reconstruction protocols. Specifically, an inverse 2D fast Fourier transform is applied to reconstruct individual single-coil images, which are subsequently combined into the final target image using pixel-wise RSS aggregation:
\begin{equation}
  \mathbf{Y} = \sqrt{\sum_{j=1}^{N} \left|\mathcal{F}^{-1}(\mathbf{k}_j)\right|^2}
  \label{eq:rss}
\end{equation}

Notably, the reference target retains $k$-space noise, mirroring real clinical image characteristics. This workflow yields a paired training sample $(\mathbf{X}, \mathbf{Y})$, where $\mathbf{X}$ represents the zero-filled, undersampled multi-coil $k$-space input and $\mathbf{Y}$ denotes the fully sampled reference target used for loss computation. Although the structural content of $(\mathbf{X}, \mathbf{Y})$ stems from natural images, the physical mapping between the input and target strictly obeys multi-coil MRI forward physics—establishing physics-simulated data as a viable and scalable alternative to clinical datasets for training generalizable reconstruction models.

Simulated data can be used either exclusively for training or combined with real clinical acquisitions. When mixing datasets, real multi-coil measurements are aligned to match the spatial and channel dimensions of the simulated data: coil channels are zero-padded if fewer than the target dimension or truncated if greater. Simulated and real samples are blended at the batch level during training.

\subsection{Model architecture}\label{sec_model}
Traditional MRI reconstruction models focus on embedding domain-specific priors (for example, explicit sensitivity map estimation and data consistency modules) into increasingly complex, task-tailored architectures to mitigate data scarcity\cite{E2EVarNet,HUMUS-Net,PDAC,PromptMR+,SDUM}. In contrast, we argue that the virtually unlimited dataset scale provided by physics simulation necessitates a fundamental rethink of model design. We center our architecture on capacity and scalability while minimizing domain-specific architectural constraints, to absorb the vast diversity of physics-simulated data and continuously improve with compute scaling.

In SIM, we introduce the multi-channel Vision Transformer (mcViT), which embeds two simple MRI principles into a generic Vision Transformer architecture: (1) applying an inverse Fourier transform to zero-filled, undersampled $k$-space yields a spatial representation distorted by aliasing artifacts; and (2) if individual coil images are accurately recovered, the composite target image is obtained via pixel-wise RSS combination. Based on the first principle, the model receives aliased images as input and predicts the aliasing residual rather than directly estimating the clean image, establishing an outer residual connection. Guided by the second principle, the backbone reconstructs each single-coil image independently before applying a terminal RSS combination operator to produce the final output $Z$, formulated as below:
\begin{equation}
  \mathbf{Z} = \sqrt{\sum_{i = 1}^{N} |\mathbf{x}_i + \mathcal{D}_{\theta}(\mathbf{x}_i)|^2}
  \label{eq:model}
\end{equation}
where $\mathbf{x}_i = \mathcal{F}^{-1}(\tilde{\mathbf{k}}_i) \in \mathbb{R}^{2 \times H \times W}$ represents the aliased input image for the $i$-th coil, with the two-channel lead dimension concatenating the real and imaginary components of the complex-valued MR image. $H$ and $W$ represent the readout and phase-encoding grid dimensions in $k$-space, corresponding to the image-space height and width under Cartesian sampling. $\mathcal{D}_{\theta}$ denotes a Vision Transformer-based learnable dealiasing module parameterised by $\theta$.

Here, $\mathcal{D}_{\theta}$ is adapted from the Diffusion Transformer architecture (DiT)\cite{DiT}, which maps image-to-image representations by projecting the input image into a sequence of patch tokens via an initial linear layer, processing these tokens through stacked self-attention layers, and reconstructing the spatial image via a final linear projection. We incorporate 2D Rotary Position Embeddings (RoPE)\cite{rope} instead of DiT's default additive sinusoidal positional embeddings. To improve computational efficiency, particularly for large models operating with fine-grained patch sizes, we replace full self-attention with an alternating mechanism that interchanges dense global self-attention with 1D sparse attention layers\cite{ueps}.

\subsection{Scalable training}\label{sec_train}
Model parameters for the self-attention layers are initialized following CLIP\cite{CLIP}, while the final linear layer's weights and biases are initialized to zero following DiT\cite{DiT}. Supervised training optimizes a Mean Absolute Error (MAE) loss between predicted reconstructions and ground-truth targets derived from simulated data pairs. Optimization is performed using AdamW\cite{adamw} with hyperparameters $\beta_1 = 0.9$, $\beta_2 = 0.95$, a weight decay of $0.1$, and a total batch size of $24$. The learning rate follows a multi-stage schedule similar to that used in DeepSeek\cite{deepseekv3}: linear warmup ($1 \times 10^{-6}$ to $1 \times 10^{-4}$ over the first $1\%$ of iterations), constant hold at $1 \times 10^{-4}$ through $20\%$ of training, cosine decay to $5 \times 10^{-6}$ from $20\%$ to $95\%$ of steps, and a constant floor of $5 \times 10^{-6}$ for the final $5\%$ of training. Gradient norms are clipped at a maximum threshold of $1.0$.

Our co-design of physics-simulated training data and high-capacity architecture enables highly scalable training. Reconstruction performance scales continuously with increased model parameters ($N$), finer patch size ($P$), expanding simulated dataset size ($D$), increasing coil count of simulated dataset ($C$), and extended training steps ($S$). In contrast to traditional clinical-data-trained models that suffer early convergence due to data scarcity, our paradigm shows no performance plateau, continuously yielding gains as compute scales. Consequently, the configurations reported in this work represent performance checkpoints constrained only by allocated training budgets.

We train and evaluate SIM across diverse computational budgets:

\begin{itemize}
    \item \textbf{Ablation Studies (Minimal Compute Budget):} \\
    Models are trained strictly on simulated data with parameter configurations set to $N = 202\text{M}$, patch size $P = 10$, simulated dataset size $D_{\text{sim}} = 0.45\text{M}$, coil count $C = 8$, epochs $E = 8$, and effective acceleration factor $R = 3.2$. 

    \item \textbf{Benchmarking \& Cross-Protocol Evaluation (Modest Compute Budget):} \\
    The benchmark model utilizes $N = 510\text{M}$ parameters, $P = 5$, $C = 8$, $E = 8$, and $R = 3.2$, co-trained on $D_{\text{sim}} = 1.01\text{M}$ simulated samples and $D_{\text{real}} = 70,748$ real clinical samples.

    \item \textbf{Interpretability Study (Modest Compute Budget):} \\
    The model is instantiated with $N = 510\text{M}$, $P = 5$, $D_{\text{sim}} = 1.01\text{M}$, $D_{\text{real}} = 70,748$, $C = 8$, $E = 7$, and $R \in \{1.9, 3.2, 4.2, 5.1\}$.

    \item \textbf{Downstream \& Clinical Studies (Maximal Compute Budget):} \\
    Our primary high-capacity model is trained with $N = 1.01\text{B}$ parameters, $P = 5$, $D_{\text{sim}} = 1.01\text{M}$, $D_{\text{real}} = 70,748$, $C = 8$, $E = 7$, and $R \in \{1.9, 3.2, 4.2, 5.1\}$.

    \item \textbf{Systematic Scaling Experiments:}
    \begin{itemize}
        \item \textbf{Joint Data \& Model Scaling:} Model capacity spans $N \in \{11\text{M}, 25\text{M}, 49\text{M}, 85\text{M}, 135\text{M}, 202\text{M}, 567\text{M}\}$ with $P = 10$ and $R = 3.2$. For experiments utilizing simulated data, data augmentation was disabled, with dataset sizes set to $D_{\text{sim}} \in \{1\text{k}, 2\text{k}, 4\text{k}, 10\text{k}, 20\text{k}, 50\text{k}\}$ with $C = 8$. For real data based experiments, $D_{\text{real}} \in \{10\text{k}, 20\text{k}, 50\text{k}, 71\text{k}\}$ with $C = 20$. Epochs ($E$) are dynamically set to ensure total sample exposure $E \times D \approx 3.6 \times 10^6$.

        \item \textbf{Model Size Scaling:} Parameter size scales across $N \in \{11\text{M}, 25\text{M}, 49\text{M}, 85\text{M}, 135\text{M}, 202\text{M}, 567\text{M}\}$ with fixed hyper-parameters $P = 10$, $D_{\text{sim}} = 1.12\text{M}$, $C = 15$, $E = 8$, and $R = 3.2$.

        \item \textbf{Patch Size Scaling:} Patch size is varied across $P \in \{5, 8, 10, 16, 20, 32\}$ with $N = 102\text{M}$, $D = 0.89\text{M}$, $C = 8$, $E = 10$, and $R = 3.2$.
    \end{itemize}
\end{itemize}

Models were trained using PyTorch distributed training\cite{pytorch} on single-node servers equipped with eight GPUs. Standard Distributed Data Parallel (DDP) was employed by default, whereas Fully Sharded Data Parallel (FSDP) was utilized when model scale exceeded single-GPU memory capacity. Experiments were partitioned across two hardware nodes housing eight NVIDIA RTX 4090 GPUs and eight NVIDIA RTX 5090 GPUs, respectively. The computational framework was implemented in Python (v3.12.12) using PyTorch (v2.10.0) and CUDA (v13.0). To optimize throughput during training and evaluation, we leveraged \texttt{torch.compile} alongside Automatic Mixed Precision (AMP) using the \texttt{bfloat16} floating-point format. AMP, combined with gradient accumulation and activation checkpointing, also mitigated memory bottlenecks, enabling the execution of large-capacity models within constrained CUDA memory bounds. Unless otherwise specified, all runtime measurements were benchmarked on a single NVIDIA RTX 5090 GPU.

\subsection{Evaluation protocols}\label{sec_eval}
\paragraph{Benchmark dataset assembly}\label{sec_assem}
To evaluate model generalization across diverse imaging conditions, we assembled a comprehensive benchmark comprising 14 multi-coil complex $k$-space datasets. To establish reproducible and quantitative comparisons, eight public datasets were integrated: fastMRI Brain\cite{fastMRI}, fastMRI Knee\cite{fastMRI}, CC-359\cite{ccbrain}, M4Raw\cite{m4raw}, M4Raw GRE\cite{m4raw}, AHEAD\cite{AHEAD}, Stanford 2D\cite{stanford2d}, and Stanford 3D\cite{stanford3d}. To further challenge generalization across broader imaging protocols and contrast mechanisms, six in-house datasets were incorporated: MULTIPLEX (quantitative MRI), PCrCEST (dynamic molecular MRI), STAR (routine brain protocols in healthy controls and Alzheimer’s disease cohorts for radiologist evaluation), Multi-Anatomy (broad anatomical coverage), Multi-Contrast (broad clinical contrast range), and DCE (dynamic liver perfusion and contrast agent kinetics). Extended Data Table~\ref{tab_dataset} details the key parameters of each dataset—including vendor, acquisition site, magnetic field strength ($B_0$), receiver coil count, anatomical coverage, slice orientation, pulse sequence type, contrast mechanism, and total counts (subjects, volumes, and slices)—spanning substantial distribution shifts across hardware, populations, anatomies, and protocols.

To eliminate selection bias and ensure fully reproducible benchmarking, evaluation was conducted on entire contiguous volumes without arbitrary slice selection. Except for fastMRI Brain, all benchmark datasets served exclusively for OOD evaluation. Specifically, the six in-house cohorts and four public datasets (CC-359, M4Raw GRE, Stanford 2D, and Stanford 3D) were designated entirely as test data. For M4Raw, the official test split was evaluated in full. For fastMRI Knee—which publicly provides only training and validation splits—we partitioned the official validation set and reserved the final 150 volumes for OOD testing. For AHEAD, a representative subset of the final 20 volumes (comprising 5 echoes per volume, totaling 100 evaluation volumes) was evaluated to maintain computational tractability across all baseline evaluations. Peripheral background slices at the boundaries of high-resolution 3D volumes were excluded solely due to lack of signal; exact retained slice counts and volume specifications are detailed in Extended Data Table~\ref{tab_dataset}.

\paragraph{Data harmonization and preprocessing}\label{sec_prepro}
Because our study focused on 2D multi-slice MRI—the predominant clinical acquisition paradigm—all 3D volume acquisitions were converted into equivalent 2D multi-slice formats by applying a one-dimensional inverse Fourier transform along the partition axis, yielding a stack of 2D $k$-space slices. Certain volumetric datasets were reformatted into distinct evaluation test sets by acquisition plane, yielding separate axial and sagittal test sets for CC-359, and coronal and sagittal test sets for Stanford 3D. Other volumetric acquisitions (AHEAD, MULTIPLEX, STAR 3D, Multi-Contrast, and DCE datasets) were reformatted along the axial plane. For multi-echo or dynamic datasets, individual echoes (AHEAD and MULTIPLEX) and distinct temporal frames (PCrCEST and DCE) were evaluated as independent target volumes.

Certain datasets were originally acquired under conservative acceleration to balance image quality against scan time constraints and subject motion. These included the public AHEAD dataset ($2\times$)\cite{AHEAD} and in-house cohorts: STAR 3D ($2.6\times$), STAR 2D ($2.2\times$), Multi-Contrast ($5.3\times$), and DCE ($2.5\times$). To enable controlled retrospective evaluation at target acceleration factors, pseudo-fully-sampled multi-coil $k$-space data were generated for the natively undersampled in-house datasets by applying L1-ESPIRiT\cite{ESPIRiT} (via the Berkeley Advanced Reconstruction Toolbox (BART)\cite{bart}) to recover complex coil-wise images and performing Fourier transform on each channel. Pseudo-fully-sampled $k$-space data generation for AHEAD followed the pipeline described in its original publication\cite{AHEAD}.

To reconcile matrix size variations across datasets and ensure compatibility across evaluated baselines, all $k$-space data were standardized to a $320 \times 320$ matrix size following fastMRI conventions\cite{fastMRI}. Datasets substantially exceeding this footprint (e.g., Multi-Anatomy dataset) underwent isotropic central $k$-space cropping bounded by a maximum dimension of 320, while remaining datasets were adjusted via image-space cropping or padding following established protocols\cite{HUMUS-Net,PDAC}. $k$-Space zero-padding was omitted to avoid artificially dense sampling in the true $k$-space center, which falsely inflates reconstruction performance\cite{datacrime}. If image-space padding was applied, the padded region was removed prior to quantitative evaluation. To standardize multi-coil array dimensions for batch processing within each dataset, zero-padding was appended along the coil axis up to the maximum coil count of that dataset without modifying underlying physical data. Finally, intensity normalization was performed by dividing each volume by its global maximum.

\paragraph{Quantitative evaluation metrics}\label{sec_metric}
Objective reconstruction fidelity was evaluated at the 2D slice level using peak signal-to-noise ratio (PSNR) and structural similarity index measure (SSIM), computed against fully sampled references derived from RSS multi-coil combination. For a reconstructed slice $\mathbf{Z}$ and ground-truth target slice $\mathbf{Y}$, PSNR is defined as:
\begin{equation}
  \operatorname{PSNR}(\mathbf{Z}, \mathbf{Y}) = 10 \log_{10} \left( \frac{\operatorname{MAX}(\mathbf{Y})^2}{\operatorname{MSE}(\mathbf{Z}, \mathbf{Y})} \right)
  \label{eq:psnr}
\end{equation}
where $\operatorname{MSE}(\mathbf{Z}, \mathbf{Y}) = \frac{1}{HW} \sum_{i=1}^{H} \sum_{j=1}^{W} \left( \mathbf{Z}_{i,j} - \mathbf{Y}_{i,j} \right)^2$ represents the mean squared error across an $H \times W$ spatial matrix, and $\operatorname{MAX}(\mathbf{Y})$ denotes the maximum target intensity calculated globally across the entire 3D volume, following fastMRI protocols\cite{fastMRI}.

SSIM was computed using scikit-image (v0.26.0) according to:
\begin{equation}
  \operatorname{SSIM}(\mathbf{Z}, \mathbf{Y}) = \frac{(2\mu_{\mathbf{Z}} \mu_{\mathbf{Y}} + C_1)(2\sigma_{\mathbf{Z}\mathbf{Y}} + C_2)}{(\mu_{\mathbf{Z}}^2 + \mu_{\mathbf{Y}}^2 + C_1)(\sigma_{\mathbf{Z}}^2 + \sigma_{\mathbf{Y}}^2 + C_2)}
  \label{eq:ssim}
\end{equation}
where $\mu_{\mathbf{Z}}$ and $\mu_{\mathbf{Y}}$ denote the local means, $\sigma_{\mathbf{Z}}^2$ and $\sigma_{\mathbf{Y}}^2$ represent the local variances, and $\sigma_{\mathbf{Z}\mathbf{Y}}$ is the local covariance of $\mathbf{Z}$ and $\mathbf{Y}$. Small constants $C_1 = (K_1 L)^2$ and $C_2 = (K_2 L)^2$ stabilize the division, where the dynamic data range $L$ is set to the global 3D volume maximum, analogous to PSNR.

\paragraph{Qualitative radiologist ratings}\label{sec_rating}
To evaluate the clinical fidelity and diagnostic utility of the reconstructed MRI volumes, a qualitative reader study was conducted by two board-certified radiologists, each possessing over 10 years of clinical experience. Evaluators were fully blinded to reconstruction methods, acceleration factors, and subject identifiers. For each evaluation session, reconstructed $T_1$, $T_2$, and FLAIR volumes corresponding to the same subject, reconstruction method, and acceleration factor were presented simultaneously alongside an explicitly labeled fully sampled reference scan to assist in differentiating true anatomical structures from artifacts. Crucially, fully sampled references were also anonymized, randomized, and included directly within the candidate method pool for scoring; this served to enable direct statistical comparisons against the gold standard while validating whether readers consistently assigned top ratings to the reference volumes. All evaluations were performed independently on clinical diagnostic workstations across two standardized rating scales:

Image quality was evaluated on a 5-point Likert scale defined as follows:
\begin{itemize}
    \item \textbf{5 (Excellent):} Superior image quality with sharp structural depiction, high SNR, and no detectable reconstruction artifacts.
    \item \textbf{4 (Good):} High image quality with minor, non-diagnostic artifacts or subtle blurring that does not compromise fine anatomical structures.
    \item \textbf{3 (Fair):} Moderate image quality with noticeable artifacts or noise; anatomical structures remain recognizable, but fine details are partially degraded.
    \item \textbf{2 (Poor):} Severe image quality degradation due to prominent artifacts, blur, or motion distortion, substantially compromising anatomical interpretation.
    \item \textbf{1 (Unacceptable):} Non-evaluable image quality characterized by severe structural distortion, loss of diagnostic features, or pervasive reconstruction failure.
\end{itemize}

Clinical usability was evaluated using a 3-point categorical scale tailored for decision-making utility:
\begin{itemize}
    \item \textbf{Diagnostic:} Image quality fully supports definitive primary diagnosis and fine morphological characterization without reservation.
    \item \textbf{Screening-only:} Image quality is sufficient to identify gross pathology, spatial anomalies, or major lesions, but insufficient for detailed characterization or definitive diagnosis.
    \item \textbf{Non-diagnostic:} Severe image degradation precludes reliable clinical assessment or lesion identification, requiring re-scan acquisition.
\end{itemize}

\subsection{Baselines}\label{sec_bl}
We benchmarked our approach against representative state-of-the-art methods spanning the primary paradigms in MRI reconstruction:

\begin{itemize}
    \item \textbf{Supervised methods:} Fully supervised baselines—including the end-to-end UNet\cite{Unet} and deep unrolled architectures (E2EVarNet\cite{E2EVarNet}, PDAC\cite{PDAC}, and NoVar\cite{NoVar})—were retrained using their official codebases. Models were trained strictly according to their recommended configurations, while matching our training dataset (the entire fastMRI Brain training set\cite{fastMRI}) and undersampling mask patterns to ensure a fair comparison.

    \item \textbf{Generative method:} For the diffusion-based DDS\cite{DDS}, we directly evaluated the publicly released model weights following its official evaluation protocol, which uses Gaussian undersampling masks and multi-coil target combination with estimated sensitivity maps. When evaluated under our primary setting—using equispaced undersampling masks and RSS target combination—DDS exhibited severe performance degradation; hence, its native evaluation setup was retained.

    \item \textbf{Untrained methods:} Scan-specific, test-time optimization baselines—including AeSPa (self-supervised)\cite{Aespa}, IMJENSE (implicit neural representation)\cite{IMJENSE}, and SCAMPI (deep image prior)\cite{SCAMPI}—were applied directly to individual test scans following their officially released codebases and default hyperparameter settings.

    \item \textbf{Compressed sensing method:} L1-ESPIRiT\cite{ESPIRiT}, as implemented in BART\cite{bart}, was applied via scan-specific optimization using default parameters.
\end{itemize}

All baseline methods and SIM were evaluated on the complete test sets, except for untrained approaches (AeSPa, IMJENSE, SCAMPI), which were restricted to a 500-slice subset of each test set due to their substantial computational overhead.

\subsection{Ablation study}\label{sec_ablation}
To evaluate the individual contributions of the core pillars within our co-design framework—simulated training data, model architecture, and learning objective—we conducted a series of ablation studies (Fig.~\ref{fig_simudata}c; baseline configuration detailed in Section~\ref{sec_train}). Our findings reveal that optimal design choices shift significantly when transitioning from conventional small-data regimes to large-scale data settings. While established approaches such as the E2EVarNet architecture and structural similarity (SSIM) loss excel in data-scarce regimes\cite{E2EVarNet}, they underperform relative to the mcViT model and mean absolute error (MAE) loss when trained on large-scale simulated data. This shift underscores the necessity of a scalability-centered co-design, contrasting with traditional workflows that decouple data simulation from model architecture design without joint scaling. Furthermore, ablating core simulated data components, such as magnitude alignment, noise-corrupted targets, and random coil sensitivities, induces a substantially larger performance drop than architectural ablations. This demonstrates that closing the physics-based simulation-to-real gap is foundational to achieving generalizable, high-performance MRI reconstruction.

\subsection{Quantitative MRI study}\label{sec_qmridetail}
Multi-parametric MRI protocols acquire multi-echo sequences with varying contrast weighting to enable spatial mapping of fundamental physical properties, such as longitudinal $T_1$ and transverse $T_2/T_2^*$ relaxation times. To evaluate the zero-shot generalizability of SIM on multi-echo acquisitions, we perform brain parameter mapping experiments across two independent cohorts acquired with distinct multi-parametric sequences:

\begin{enumerate}
    \item \textbf{AHEAD Dataset\cite{AHEAD}:} A public benchmark acquired at the Spinoza Centre for Neuroimaging (Amsterdam, Netherlands) using a 7~T scanner (Philips Achieva) with an 3D MP2RAGEME sequence ($N = 20$ subjects)\cite{mp2rageme}.

    \item \textbf{MULTIPLEX Dataset:} An in-house cohort acquired at our institution (Shenzhen, China) using a 3~T scanner (UIH uMR 790) with a 3D MULTIPLEX sequence ($N = 20$ subjects)\cite{mtp}.
\end{enumerate}

\noindent Both acquisition protocols support multi-parametric quantification; here, we focus on $T_2^*$ mapping as a representative task. For the AHEAD dataset, four echoes from the second inversion block were selected, corresponding to echo times (TEs) of $3.0$, $11.5$, $19.0$, and $28.5\,\text{ms}$. For the MULTIPLEX dataset, the final five echoes acquired from the first flip-angle group were selected, with TEs of $3.75$, $7.55$, $11.35$, $15.15$, and $18.95\,\text{ms}$.

SIM and the comparison baseline model (E2EVarNet) are applied independently to each multi-echo 3D volume. Voxel-wise parameter estimation is then performed separately on the fully sampled reference images, the SIM reconstructions, and the E2EVarNet reconstructions, with the reference-derived maps serving as the ground-truth $T_2^*$ reference. Specifically, quantitative $T_2^*$ estimation is conducted via non-linear least-squares optimization using a mono-exponential signal decay model:
\begin{equation}
S(\mathrm{TE}) = S_0 \exp\left(-\frac{\mathrm{TE}}{T_2^*}\right) + C
\label{eq:t2starfit}
\end{equation}
where $S(\mathrm{TE})$ denotes the magnitude signal intensity at echo time $\mathrm{TE}$, $S_0$ is the initial signal amplitude, $T_2^*$ is the effective transverse relaxation time, and $C$ accounts for a constant baseline noise offset. The resulting voxel-wise $T_2^*$ estimates are subsequently reshaped into 2D quantitative parameter maps.

Non-brain structures and background were removed using ANTsPy (v0.6.3) to restrict quantitative evaluation to brain tissue. Brain masks extracted from the first echo of the fully sampled reference images were applied directly to all reconstructed $T_2^*$ maps to enforce a consistent evaluation region of interest (ROI). Accuracy of the reconstructed $T_2^*$ maps was quantified at the 2D slice level using Mean Absolute Error (MAE), expressed in milliseconds ($\text{ms}$).

\subsection{Visualization}\label{sec_vis}
\paragraph{Data visualization}\label{sec_datavis}
Although our physics-simulated training images are derived from natural scenes and lack global anatomical context, they embody diverse local contrast patterns relevant to clinical imaging. To evaluate local similarity between simulated and clinical distributions, patch-level features were mapped into a shared representation space. Target images across simulated and real cohorts were partitioned into $14 \times 14$-pixel patches. For real data, a two-patch-wide peripheral margin was excluded to avoid non-informative background regions. Using a pretrained DINOv2 ViT-B/14 model with register tokens\cite{dinov2}, output embeddings were harvested from the 8th layer to yield an 768-dimensional feature vector per patch, which was subsequently $L_2$-normalized to unit length.

Dimensionality reduction was fitted exclusively on real-data features to project both simulated and real patches into a shared, real-anchored representation space. Principal Component Analysis (PCA) was first fitted to compress the feature dimension from 768 to 64. Linear Discriminant Analysis (LDA)\cite{LDA} was subsequently applied using dataset provenance as class labels, to identify the top two primary discriminant axes.

Simulated data were compared against five representative clinical datasets: fastMRI Brain, fastMRI Knee, CC-359 Axial, M4Raw GRE, and AHEAD Echo3 (Extended Data Table~\ref{tab_dataset}). Transformations were fit on 4 patches sampled without replacement per image across 512 randomly selected slices per real dataset and 2,048 simulated images. For display, 384 patch features per clinical cohort and 4,096 patch features from simulated images were randomly sub-sampled and plotted along the first two discriminant axes, color-coded by dataset source (Fig.~\ref{fig_simudata}b).

\paragraph{Model visualization}\label{sec_modelvis}
To interpret the spatial attention mechanisms learned by the model, query-specific attention weights are extracted during inference. For a given receiver coil $c$, transformer layer $l$, and attention head $h$, the scalar attention weight $A_{i,j}^{c,l,h}$, which quantifies the influence of the $j$-th key patch on the $i$-th query patch, is computed as:
\begin{equation}
A_{i,j}^{c,l,h} = \operatorname{softmax}_j \left( \frac{\mathbf{q}_i^{c,l,h} (\mathbf{k}_j^{c,l,h})^\top}{\sqrt{d}} \right)
\label{eq:attnbasic}
\end{equation}
where $\mathbf{q}_i^{c,l,h} \in \mathbb{R}^{d}$ and $\mathbf{k}_j^{c,l,h} \in \mathbb{R}^{d}$ denote the $i$-th query vector and $j$-th key vector, respectively, $d$ represents the feature dimension per attention head, and $\operatorname{softmax}_j$ denotes softmax normalization across all key patches, with the subscript $j$ indicating the resulting $j$-th component.

To analyze layer- and head-specific attention behaviors independently of receiver coil variations, attention weights are aggregated across all $C$ coils. The resulting coil-averaged attention weight at the $l$-th layer and $h$-th attention head, $\tilde{A}_{i,j}^{l,h}$, is formulated as:
\begin{equation}
\tilde{A}_{i,j}^{l,h} = \frac{1}{C} \sum_{c=1}^{C} A_{i,j}^{c,l,h}
\label{eq:attncoilavg}
\end{equation}

Furthermore, a global attention map $\bar{A}_{i,j}$ is constructed by averaging across all coils, layers, and attention heads:
\begin{equation}
\bar{A}_{i,j} = \frac{1}{C L H} \sum_{c=1}^{C} \sum_{l=1}^{L} \sum_{h=1}^{H} A_{i,j}^{c,l,h}
\label{eq:attnavg}
\end{equation}
where $L$ and $H$ denote the total number of transformer layers and attention heads, respectively.

For spatial visualization of a specific query patch $i$, the corresponding key-wise attention distribution across all spatial patches $j$ is reshaped onto the 2D spatial patch grid, min-max normalized to $[0,1]$, and overlaid onto the aliased input image.

\subsection{Statistical analysis}\label{sec_stats}
Quantitative metrics across dataset slices are reported as $\text{mean} \pm \text{s.d.}$ Statistical significance ($P$ values) for continuous reconstruction metrics (e.g., PSNR and SSIM) between paired model outputs was evaluated using two-sided paired Student’s $t$-tests based on volume-averaged metric values paired per subject. Discrete reader study ratings—comprising image quality (1–5 Likert scale) and diagnostic usability (1–3 scale) independently assigned by two radiologists—were compared between paired methods using a two-sided clustered Wilcoxon signed-rank test. Each comparison comprised 100 paired observations derived from 50 subject scans evaluated by two readers. To account for potential intra-subject correlation between reader evaluations of the same scan, individual subjects were defined as clustering units (50 clusters, two observations per cluster). Inter-reader agreement was quantified using quadratically weighted Cohen’s $\kappa$ for image quality scores and linearly weighted Cohen’s $\kappa$ for diagnostic usability. Statistical significance was established at $P < 0.05$. In figures, $P$ values are designated as follows: ${}^{*}P < 0.05$, ${}^{**}P < 0.01$, ${}^{***}P < 0.001$, ${}^{****}P < 0.0001$. Sample sizes were determined by dataset availability and are detailed in Extended Data Table~\ref{tab_dataset}. No data were excluded from quantitative evaluations. All statistical computations were executed in Python (v3.12.12) using SciPy (v1.15.3) and scikit-learn (v1.9.0).

\subsection{Ethics approval}\label{sec_ethics}
This study analyzed both publicly available and in-house clinical datasets collected across multiple institutions. Publicly available benchmark datasets were acquired from their respective open-source repositories under their original ethical clearances and data-use agreements. For the in-house clinical cohorts, ethical approvals were granted by the corresponding institutional governing bodies:
\begin{itemize}
    \item \textbf{Shenzhen Institutes of Advanced Technology:} The study protocol was approved by the Institutional Review Board of Shenzhen Institutes of Advanced Technology, Chinese Academy of Sciences (Approval Nos. SIAT-IRB-221115-H0631 and SIAT-IRB-200315-H0454). Owing to the fully anonymized, retrospective design of this analysis, the requirement for written informed consent was waived by the Institutional Review Board.
    \item \textbf{Peking University Shenzhen Hospital:} The study protocol was approved by the Ethics Committee of Peking University Shenzhen Hospital (Approval No. 2022-160-01). Written informed consent was obtained from each participant before enrolment.
\end{itemize}
The study adheres to the principles of the Declaration of Helsinki. All clinical datasets were de-identified prior to transfer to the study investigators.




\backmatter





\clearpage

\begin{appendices}

\renewcommand{\theHfigure}{A\arabic{figure}}

\section{Extended Data}\label{extdata}


\begin{sidewaystable}
\caption{Summary of evaluation datasets.}
\label{tab_dataset}
\setlength{\tabcolsep}{2pt}
\footnotesize
\begin{tabular*}{\textheight}{@{\extracolsep\fill}llrrrrrrrrrrr}
\toprule
\textbf{Split} & \textbf{Dataset} & \textbf{Source} & \makecell[r]{\textbf{Ana-} \\ \textbf{tomy}} & \textbf{Plane} & \textbf{Sequence} & \textbf{Contrast} & \textbf{Vendor} & \makecell[r]{\textbf{$B_0$} \\ \textbf{(T)}} & \textbf{Coil} & \makecell[r]{\textbf{Subj.} \\ \textbf{/Vol.}} & \textbf{Slice} & \textbf{Center} \\
\midrule
Train & \multirow{2}{*}{fastMRI Brain} & \multirow{2}{*}{Public\cite{fastMRI}} & \multirow{2}{*}{brain} & \multirow{2}{*}{axial} & \multirow{2}{*}{2D FSE} & \multirow{2}{1.2cm}{\makecell[r]{$T_1$/$T_2$/ \\ FL/CE}} & \multirow{2}{*}{Siemens} & \multirow{2}{*}{1.5/3} & 2–28 & 4469/4469 & 70748 & \multirow{2}{*}{NYU} \\
Test(ID) & & & & & & & & & 2–24 & 558/558 & 8852 & \\
\midrule
\multirow{16}{0.6cm}{Test (OOD)} 
& fastMRI Knee & Public\cite{fastMRI} & knee & coronal & 2D FSE & PD/PDFS & Siemens & 1.5/3 & 15 & 83/150 & 5427 & NYU \\
& CC-359 Ax & Public\cite{ccbrain} & brain & axial & 3D MPRAGE & $T_1$ & GE & 3 & 12 & 67/67 & 13802 & Calgary/Campinas \\
& CC-359 Sag & Public\cite{ccbrain} & brain & sagittal & 3D MPRAGE & $T_1$ & GE & 3 & 12 & 67/67 & 9490 & Calgary/Campinas \\
& M4Raw & Public\cite{m4raw} & brain & axial & 2D SE/FSE & $T_1$/$T_2$/FL & Xingaoyi & 0.3 & 4 & 25/400 & 7200 & Shenzhen Tech. Univ. \\
& M4Raw GRE & Public\cite{m4raw} & brain & axial & 2D GRE & $T_1$ & Xingaoyi & 0.3 & 4 & 183/366 & 6588 & Shenzhen Tech. Univ. \\
& AHEAD Echo1-5 & Public\cite{AHEAD} & brain & axial & 3D MP2RAGEME & $T_1$/$T_2^*$ & Philips & 7 & 32 & 20/100 & 16200 & Amsterdam Spinoza \\
& Stanford 2D & Public\cite{stanford2d} & mix & mix & 2D FSE & mix & GE & 3 & 3–32 & 89/89 & 2037 & Stanford Univ. \\
& Stanford 3D Cor & Public\cite{stanford3d} & knee & coronal & 3D FSE & PDFS & GE & 3 & 8 & 19/19 & 4655 & Stanford Univ. \\
& Stanford 3D Sag & Public\cite{stanford3d} & knee & sagittal & 3D FSE & PDFS & GE & 3 & 8 & 19/19 & 3249 & Stanford Univ. \\
& MULTIPLEX & In-house & brain & axial & MULTIPLEX & $T_2^*$ & UIH & 3 & 32 & 20/100 & 8400 & SIAT \\
& PCrCEST & In-house & lower leg & axial & 2D FSE & PCrCEST & UIH & 5 & 24 & 1/17 & 459 & SIAT \\
& STAR 3D & In-house & brain & axial & 3D MPRAGE & $T_1$ & UIH & 3 & 32 & 50/50 & 8000 & PKU Shenzhen Hosp. \\
& STAR 2D & In-house & brain & axial & 2D FSE & $T_2$/FL & UIH & 3 & 32 & 50/100 & 2500 & PKU Shenzhen Hosp. \\
& Multi-Anatomy & In-house & mix & mix & 2D FSE & $T_2$ & UIH & 3 & 14-42 & 1/6 & 81 & SIAT \\
& Multi-Contrast & In-house & brain/neck & axial & mix & mix & UIH & 3 & 14/50 & 2/7 & 1096 & SIAT \\
& DCE & In-house & liver & axial & 3D GRE & $T_1$ & Siemens & 3 & 15 & 1/7 & 288 & SIAT \\
\botrule
\end{tabular*}
\end{sidewaystable}

\clearpage

\begin{figure*}[h]
\centering
\includegraphics[width=1\textwidth]{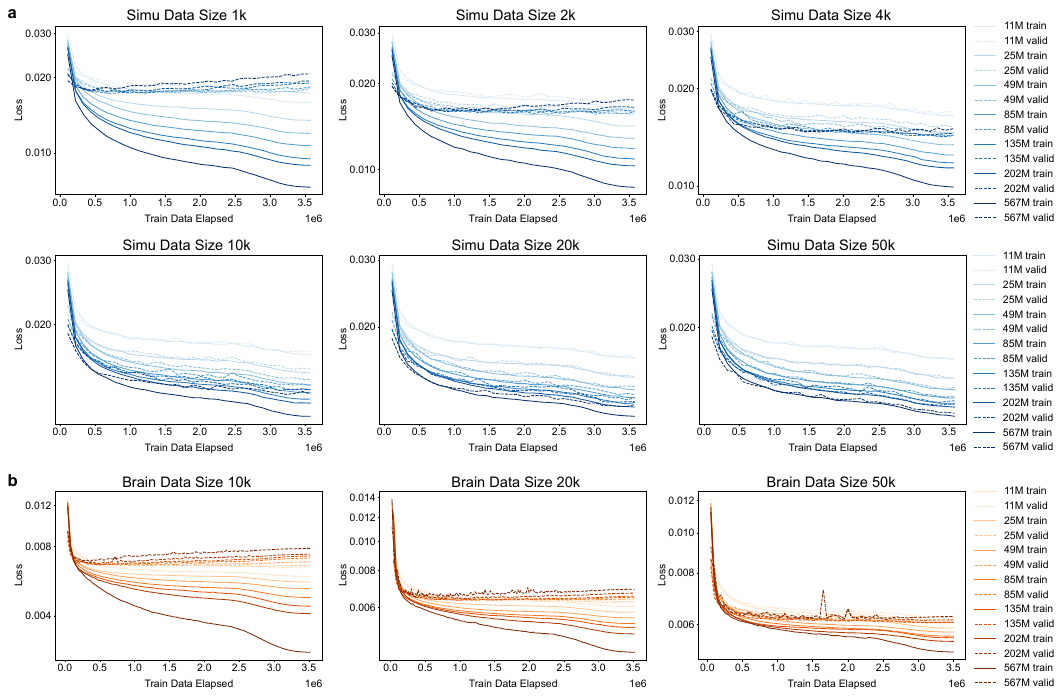}
\caption{Training Curves. a, b, Training and validation loss for a series of mcViT models trained on simulated data (a) and real fastMRI (b) across various dataset and model sizes. For both data, increasing model capacity leads to severe overfitting when data diversity is inadequate, whereas increasing dataset size reduces the training–validation gap. Scaling simulated training data to 50,000 samples prevents overfitting for models up to 567 million parameters, whereas severe overfitting persists for 50,000 real fastMRI samples, demonstrating the substantially higher diversity of simulated data at comparable dataset sizes required for continuous benefits from larger models. Valid, validation. For a and b, retrospective equispaced undersampling is applied at a $3.2\times$ effective acceleration factor.}
\label{ed_scale_loss}
\end{figure*}

\clearpage

\begin{figure}[h]
\centering
\includegraphics[width=0.8\textwidth]{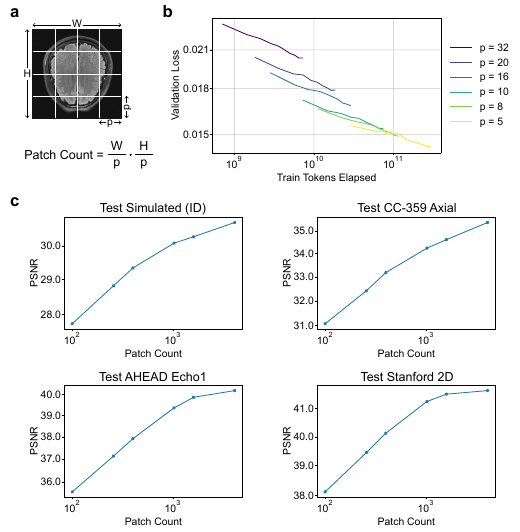}
\caption{Scaling patch size. a, Illustration of patch size ($p$) and its inverse relationship to patch count. b, Validation loss for a series of mcViT models trained on simulated data across various patch sizes, ranging from 32 to 5. Reducing patch size—thereby increasing the Transformer sequence length and computational cost—yields consistent performance gains. c, Transfer scaling. Patch size scaling behavior learned on simulated data transfers effectively to diverse real-world datasets, analogous to model size scaling. ID, in-distribution. For b and c, retrospective equispaced undersampling is applied at a $3.2\times$ effective acceleration factor.}
\label{ed_scale_ps}
\end{figure}

\clearpage

\begin{figure*}[h]
\centering
\includegraphics[width=1\textwidth]{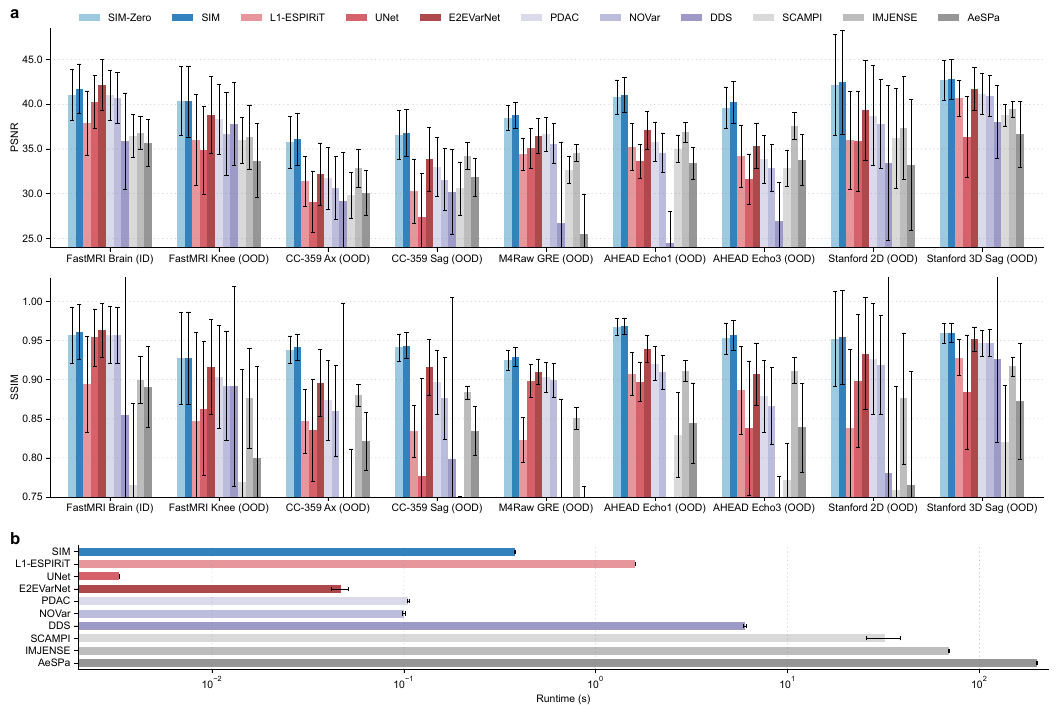}
\caption{Full quantitative comparison of zero-shot reconstruction. a, Quantitative comparison across in-distribution (ID) and out-of-distribution (OOD) benchmarks in terms of PSNR and SSIM. Both SIM and SIM-Zero (trained exclusively on simulated data) outperform all 9 competing baselines across 8 OOD benchmarks. b, Runtime comparison measured on 15-coil fastMRI knee data using a single NVIDIA RTX 5090 GPU. SIM operates via a single feed-forward inference step, achieving low latency compared to diffusion-based and untrained approaches that require multi-step iteration per sample. In a and b, error bars represent mean $\pm$ s.d., and retrospective equispaced undersampling is applied at a $3.2\times$ effective acceleration factor.}
\label{ed_bench_full}
\end{figure*}

\clearpage

\begin{figure*}[h]
\centering
\includegraphics[width=1\textwidth]{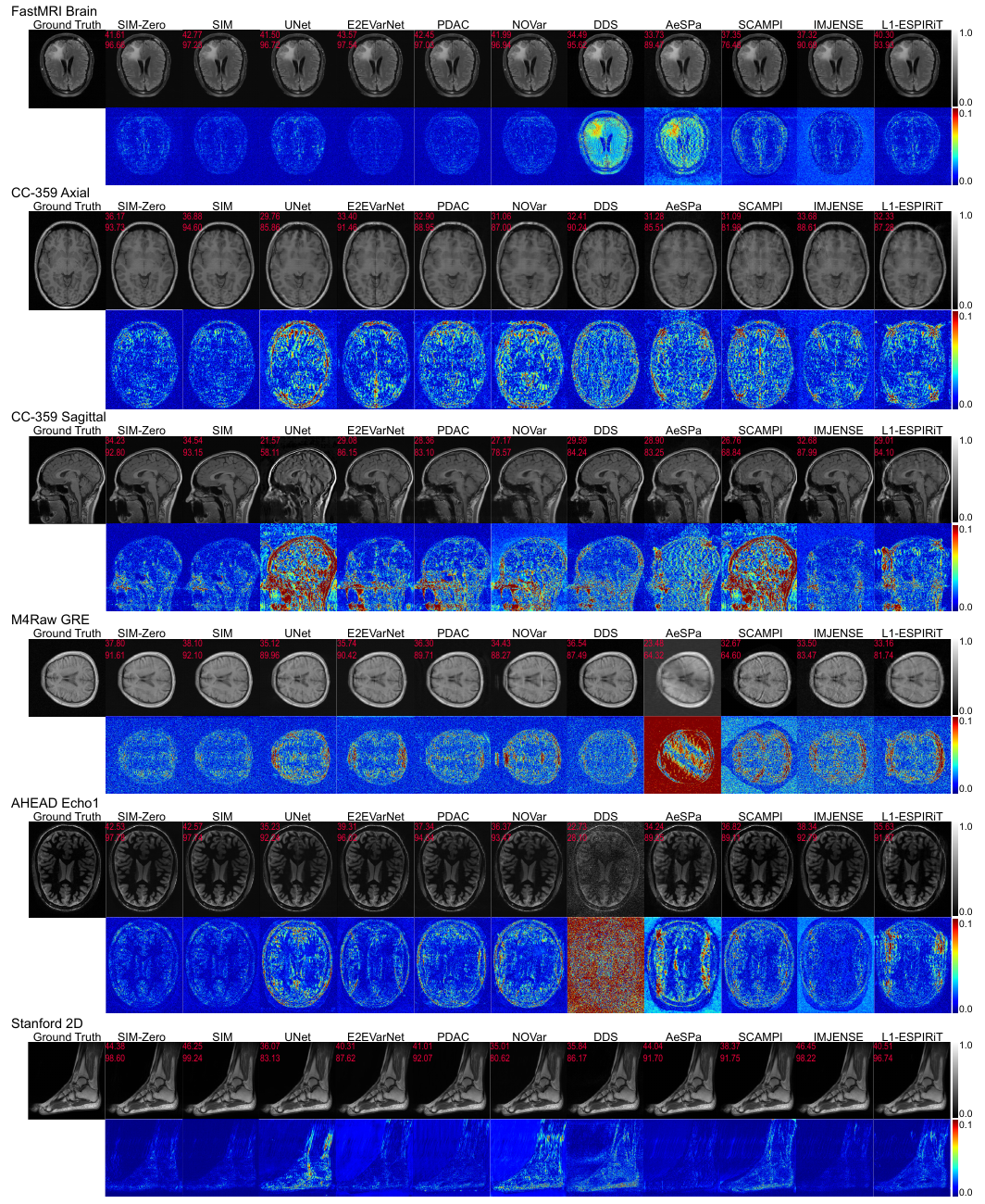}
\caption{Visual comparison of zero-shot reconstruction. Representative reconstructions from SIM, SIM-Zero (trained exclusively on simulated data), and 9 baseline methods evaluated on examples from fastMRI Brain, CC-359 Axial, CC-359 Sagittal, M4Raw GRE, AHEAD Echo1, and Stanford 2D datasets. PSNR (dB) and SSIM (\%) values for individual slices are indicated in the top-left corner of each reconstruction. Retrospective equispaced undersampling is applied at a $3.2\times$ effective acceleration factor.}
\label{ed_bench_examples}
\end{figure*}

\clearpage

\begin{figure*}[h]
\centering
\includegraphics[width=1\textwidth]{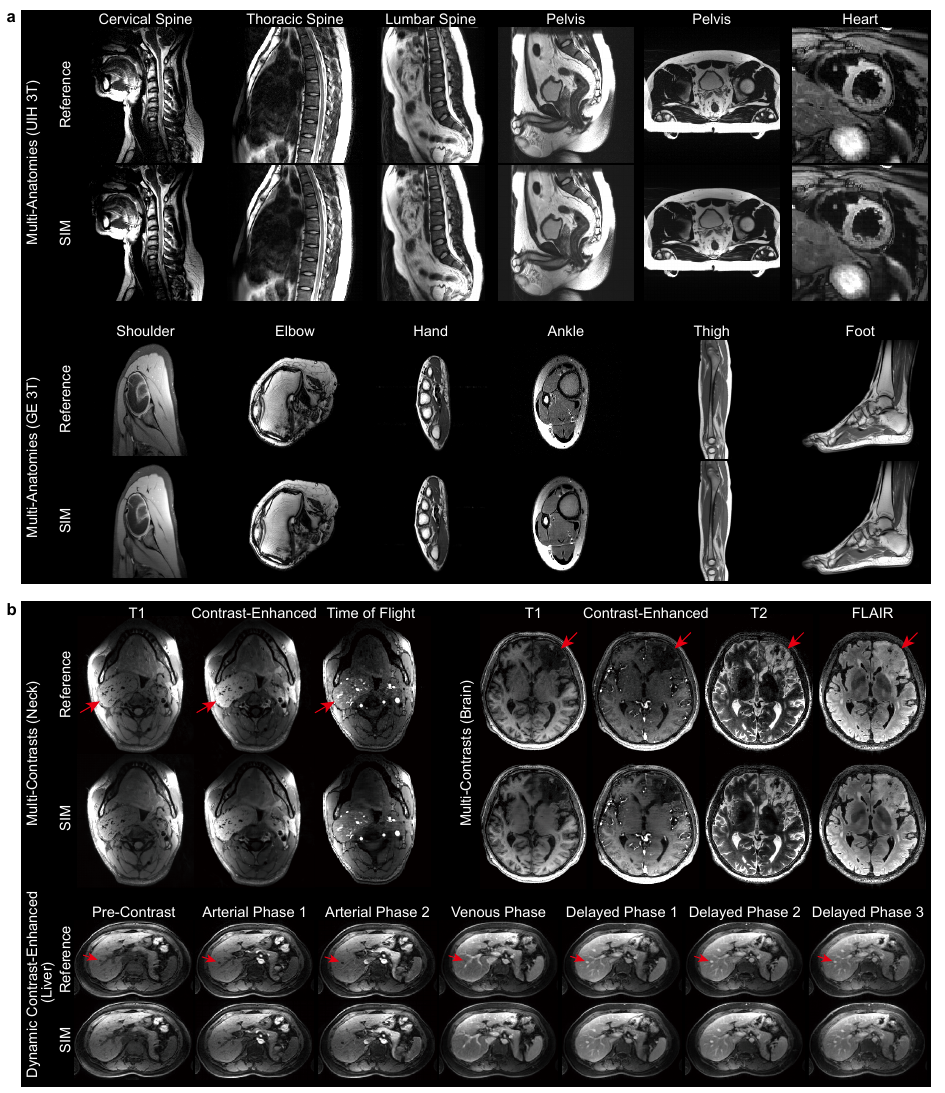}
\caption{Zero-shot reconstruction across broad anatomies and contrasts. a, Broad anatomies. Representative SIM reconstructions on an in-house cohort (Multi-Anatomy, UIH 3~T) and a public cohort (Stanford 2D, GE 3~T). The cardiac panel is cropped and enlarged around the heart region. b, Broad contrasts. Representative SIM reconstructions on in-house cohorts (Multi-Contrast, UIH 3~T; DCE, Siemens 3~T). Red arrows indicate lesions. In a and b, display windowing and leveling are matched between SIM reconstructions and reference images, and retrospective equispaced undersampling is applied at a $4.2\times$ effective acceleration factor. Details for each dataset are provided in Extended Data Table~\ref{tab_dataset}.}
\label{ed_anat_cont}
\end{figure*}

\clearpage

\begin{figure*}[h]
\centering
\includegraphics[width=1\textwidth]{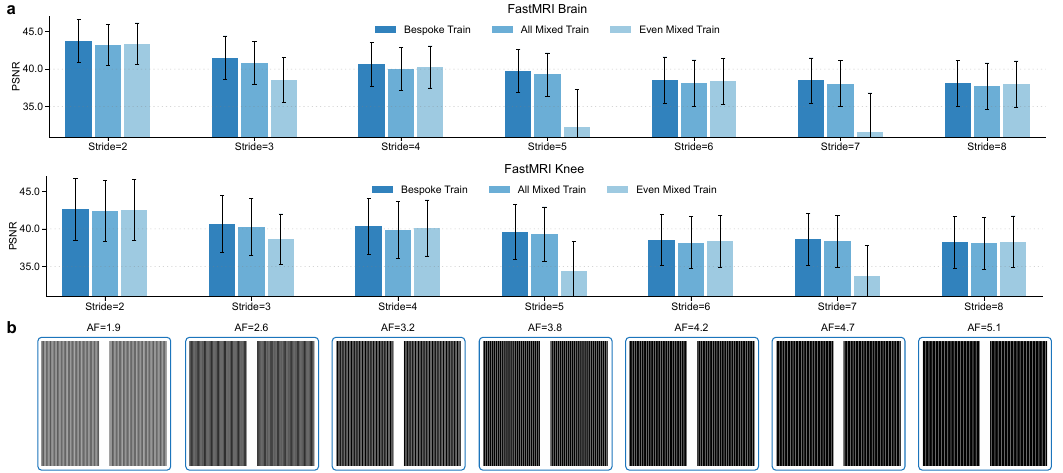}
\caption{Training strategies for multi-acceleration support. a, Quantitative comparison of three training strategies evaluated on fastMRI Brain and Knee test sets: bespoke training (single fixed acceleration factor per model), all-mixed training (randomly mixed continuous acceleration factors from 2 to 8), and even-mixed training (randomly mixed discrete acceleration factors of 2, 4, 6, and 8). Mixed-acceleration training achieves performance comparable to bespoke training provided the target acceleration factor is included during training. All training runs used simulated data to train an mcViT model under matched settings, varying only in acceleration factor configurations. Error bars represent mean $\pm$ s.d. b, Visualization of equispaced undersampling masks with an 8\% fully sampled $k$-space center and nominal acceleration factors (strides) of 2 to 8, yielding effective acceleration factors of 1.9 to 5.1. AF, acceleration factor.Training strategies for multi-acceleration support. a, Quantitative comparison of three training strategies evaluated on fastMRI Brain and Knee test sets: bespoke training (single fixed acceleration factor per model), all-mixed training (randomly mixed continuous acceleration factors from 2 to 8), and even-mixed training (randomly mixed discrete acceleration factors of 2, 4, 6, and 8). Mixed-acceleration training achieves performance comparable to bespoke training provided the target acceleration factor is included during training. All training runs used simulated data to train an mcViT model under matched settings, varying only in acceleration factor configurations. Error bars represent mean $\pm$ s.d. b, Visualization of equispaced undersampling masks with an 8\% fully sampled $k$-space center and nominal acceleration factors (strides) of 2 to 8, yielding effective acceleration factors of 1.9 to 5.1. AF, acceleration factor.}
\label{ed_multiR}
\end{figure*}



\end{appendices}


\bibliography{sn-bibliography}

\end{document}